\pdfoutput=1
\documentclass[lettersize,journal]{IEEEtran}
\usepackage{amsmath,amsfonts}
\usepackage{algorithmic}
\usepackage{algorithm}
\usepackage{array}
\usepackage[caption=false,font=normalsize,labelfont=sf,textfont=sf]{subfig}
\usepackage{textcomp}
\usepackage{stfloats}
\usepackage{url}
\usepackage{verbatim}
\usepackage{graphicx}
\usepackage{epstopdf}
\usepackage{cite}
\usepackage{bm}
\usepackage{stfloats}
\usepackage{caption}
\usepackage{subfig} 
\usepackage{array,color}
\usepackage{float}
\usepackage{amssymb}

\allowdisplaybreaks[4]

\begin{document}
	
	\title{Active RIS Aided ISAC Systems: Beamforming Design and Performance Analysis}
	
	
	
	\author{Zhiyuan Yu, Hong Ren, \IEEEmembership{Member, IEEE}, Cunhua Pan, \IEEEmembership{Senior Member, IEEE}, Gui Zhou, \IEEEmembership{Member, IEEE}, \\Boshi Wang,  Mianxiong Dong, \IEEEmembership{Member, IEEE}, and Jiangzhou Wang, \IEEEmembership{Fellow, IEEE}
		\thanks{The work of Hong Ren was supported in part by the National Natural Science Foundation of China under Grant 62101128, and in part by Basic Research Project of Jiangsu Provincial Department of Science and Technology under Grant BK20210205. The work of Cunhua Pan was supported in part by the National Natural Science Foundation of China under Grants 62201137 and 62331023. The work of Cunhua Pan and Hong Ren was supported in part by the Fundamental Research Funds for the Central Universities under Grant 2242022k60001. The work of Gui Zhou was supported by the Alexander von Humboldt Foundation.}
		
		\thanks{Z. Yu, H. Ren, C. Pan, and B. Wang are with National Mobile Communications Research Laboratory, Southeast University, Nanjing, China. (e-mail:\{zyyu, hren, cpan, boshiwang\}@seu.edu.cn). G. Zhou is with the Institute for Digital Communications, Friedrich-Alexander-University Erlangen-N\"{u}rnberg (FAU), 91054 Erlangen, Germany (email: gui.zhou@fau.de). Mianxiong Dong is with the Department of Sciences
			and Informatics, Muroran Institute of Technology, Muroran, Japan
			(e-mail: mx.dong@csse.muroran-it.ac.jp). J. Wang is with the School of Engineering, University of Kent, Canterbury CT2 7NT, U.K. (e-mail: j.z.wang@kent.ac.uk). 
			
			\emph{Corresponding authors: Hong Ren and Cunhua Pan.}}
		
	}

	
	\maketitle
	\vspace{-1cm}
	\begin{abstract}
		This paper considers an active reconfigurable intelligent surface (RIS)-aided integrated sensing and communication (ISAC) system. We aim to maximize radar signal-to-interference-plus-noise-ratio (SINR) by jointly optimizing the beamforming matrix at the dual-function radar-communication (DFRC) base station (BS) and the reflecting coefficients at the active RIS subject to the quality of service (QoS) constraints of communication users (UE) and the transmit power constraints of active RIS and DFRC BS. To tackle the optimization problem, the majorization-minimization (MM) algorithm is applied to address the nonconvex radar SINR objective function, and the resulting quartic problem is solved by developing an semidefinite relaxation (SDR)-based approach. {\color{black}Moreover, we derive the scaling order of the radar SINR with a large number of reflecting elements. Next, the transmit power allocation problem and the deployment strategy of the active RIS are studied with a moderate number of reflecting elements.} Finally, we validate the potential of the active RIS in ISAC systems compared to passive RIS. Additionally, we deliberate on several open problems that remain for future research.
	\end{abstract}
	
	
	\section{Introduction}
	As one of the key candidate technologies for the sixth generation (6G) wireless systems, integrated sensing and communication (ISAC) has received increasing   research attention in recent years. This technique can be applied to environment-aware scenarios such as vehicle-to-everything (V2X), virtual reality (VR), and augmented reality (AR). From the spectrum resource perspective, both the increase in communication capacity and the enhancement of sensing capabilities rely on larger bandwidth, thereby exacerbating the scarcity of available spectrum resources. Consequently, this imperatively prompts us to investigate strategies for achieving radar and communication coexistence (RCC)  within the same frequency band\cite{30,31}. However, additional control nodes along with additional estimation and feedback of interference channels are often required in RCC systems, which significantly increases the system design complexity\cite{29}. From the hardware deployment perspective, since radar and communication systems share similar hardware architectures,  implementing radar and communication functions on the same platform can effectively reduce the hardware overhead. In addition, through the integration of radar and communication systems, the radar gains awareness of the communication signals, enabling their utilization for sensing purposes. Thus, the dual-function radar-communication (DFRC) system is considered to be more efficient for the ISAC\cite{20}.
	
	Early contributions to ISAC primarily focused on allocating communication and sensing resources in a non-overlapping way to avoid interference, utilizing techniques such as time-division, spatial-division, and frequency-division ISAC. While these approaches are straightforward to implement in hardware, they suffer from relatively low spectrum and energy efficiencies. To address this issue, three types of fully unified ISAC waveform designs have been proposed: sensing-centric design (SCD), communication-centric design (CCD), and joint design (JD). In SCD, communication symbols are embedded into various domains of the sensing signal without causing substantial degradation in the sensing performance.  Although this approach exhibits a promising sensing performance, its applicability is restricted to scenarios that only require  low data rates. In contrast to SCD, communication performance has the highest priority in CCD. Generally, CCD performs the radar sensing function by exploiting the existing communication waveform, e.g., orthogonal frequency division multiplexing (OFDM) waveform \cite{WangjiangzhouOFDM1, WangjiangzhouOFDM2}. However, the sensing performance of the system may be significantly impaired by the randomness introduced by the communication symbols, e.g.,  cross- and auto-correlation properties. Furthermore, both CCD and SCD are constrained by the use of the existing waveforms, making it difficult to achieve a scalable balance between sensing and communication requirements. The last category, i.e., JD can potentially provide additional degrees of freedom (DoFs) for sensing and communication by conceiving an ISAC waveform from the ground up\cite{41}. Several contributions have been made in the JD to yield insightful results\cite{5,32}. Specifically, the authors of \cite{5} proposed to use the dedicated radar signal in combination with a communication signal to provide additional DoFs for sensing. Further, the optimal transmit beamforming was investigated in\cite{32}, pointing out that the specific radar signal was not required for the line-of-sight (LoS) communication links.

	{ISAC usually benefits from the availability of large bandwidths in millimeter wave (mmWave) frequency bands, creating the opportunity to achieve high data rates and improve radar range resolution. However, as the operating frequency increases, the signals become more susceptible to blockages, which can significantly degrade the sensing and communication performance.} Fortunately, reconfigurable intelligent surfaces (RISs) offer a promising solution to this problem by establishing reliable virtual links between the DFRC base station (BS) and the sensing targets. Typically, an RIS consists of a large number of passive reflecting elements, and the direction of the reflected signal can be tuned by adjusting the phase shifts of the reflecting elements. Numerous studies have demonstrated that integrating RIS into other emerging technologies can improve the signal propagation environment \cite{16,17}. {\color{black} In addition to its role as a passive reflecting element, RIS can also be utilized as a transceiver, thereby enabling more versatile and dynamic communication capabilities\cite{9324910}}.  {\color{black}There were also many studies on the integration of RIS and ISAC \cite{34,33,9782100}}. For the RIS-aided ISAC system,  two main scenarios have been typically considered \cite{2}. The first scenario is involved by using the RIS solely for the purpose of improving communication functionality, while direct transceiver-target links are used for sensing \cite{28,10,ISAC_Coal,10086570}.  In particular, \cite{28} focused on designing the transmit/receive beamforming matrices and  the phase shift of the passive RIS for multi-user scenarios, while \cite{10} aimed to minimize the transmit power of the DFRC BS by jointly designing the active and passive beamforming matrix in the presence of the interference introduced by the RIS. {\color{black} In \cite{ISAC_Coal}, the authors discussed the RIS phase shift optimization and time sharing between sensing and communication functionalities, by considering sensing time. In \cite{10086570}, the authors addressed a sum-rate maximization problem while imposing constraints on the estimation performance bound of the target's direction of arrival (DoA).} The second scenario aimed at further leveraging the benefits of RIS, particularly in enhancing radar sensing performance, by using RIS to establish a virtual link between the BS and the target. In this scenario, two primary performance indicators for sensing were considered. The contributions of \cite{2,15,27,50} optimized the beampattern at the target, which ensured that more power was illuminated to the target for sensing. In contrast, the authors of \cite{3} first considered a four-hop BS-RIS-target-RIS-BS link and investigated the received radar signal-to-interference-plus noise-ratio (SINR) at the DFRC BS. The authors of \cite{49} also considered the received four-hop echo at the BS and a Cramér-Rao bound (CRB) minimization problem was formulated. Nevertheless, due to the multiplicative fading effect, the equivalent path loss of the four-hop sensing link can be regarded as the product, rather than the sum, of the path loss of the BS-RIS, RIS-target, target-RIS, and RIS-BS links. This results in a significantly low received signal power at the BS, limiting the sensing resolution.
	
	To combat such severe signal propagation loss over the four-hop links, many new architectures of RIS have been proposed.  One such idea was to add sensors to the RIS for sensing, as suggested in \cite{39,40,42}. With this method, the sensors would perform the sensing function, thereby diminishing the number of hops and path loss between the sensors and the BS. However, implementing this approach would result in a more complex hardware design and is challenging to integrate with existing communication protocols. Another approach was to add several amplifiers\cite{6} to the passive RIS to mitigate signal attenuation through the signal amplification function, known as active RIS. The nascent technology of active RIS exhibits promising potential in mitigating the multiplicative fading issue compared to the traditional passive RIS \cite{21}. Recent contributions have shown that the active RIS yields superior performance compared to the traditional passive RIS in the communication systems with equal power budget\cite{26}. Additionally, the active RIS has garnered significant attention in the context of sensing-related applications, such as radar detection and localization \cite{35,37}. When the active RIS is used to establish a virtual link between the DFRC BS and the target, a signal that would normally experience severe fading can be amplified twice on the same active RIS panel, which we refer to multiple-self-reflection (MSR). {Most recently, there have been a few contributions that have delved into  the active RIS-aided ISAC system \cite{47,46}. Specifically, the authors of \cite{47} considered an active RIS-aided ISAC system in the scenario of cloud radio access network (C-RAN), and optimized the radar beampattern towards the sensing targets. However, the four-hop received echo was not considered in \cite{47}. The authors of \cite{46} employed the received radar SINR as a sensing metric while maximizing the secure rate of the active RIS-aided ISAC system. Nevertheless, the noise and the interference induced by the active RIS were neglected in \cite{46}. Besides, in \cite{47} and \cite{46}, the power consumption of the reflected echo at the active RIS was not considered.}
	
	Motivated by the above background, our contributions are summarized as follows:
	\begin{enumerate}
		\item In this paper, we investigate the active RIS-assisted DFRC system and analyze the received four-hop sensing signal. Specifically, we maximize the radar SINR by performing joint optimization of the transmit beamforming matrix together with the active RIS reflecting coefficients.
		Meanwhile, the quality of service (QoS) requirements of multiple users are ensured within the restrictions of the transmit power of both the BS and the active RIS. The main difficulties in solving the problem lie in the quartic nonconvex objective function and constraints. 
		
		
		\item We propose an algorithm that combines majorization-minimization (MM) with semidefinite relaxation (SDR) techniques to jointly optimize the beamforming matrix and the reflecting coefficients in an alternating manner. Specifically, the intractable objective function is firstly approximated by a bi-linear function following the MM framework. Then, the quartic objective function and constraints are transformed into quadratic expressions of the auxiliary variable and the SDR method is utilized to solve this problem. Besides, the initial point of the alternative optimization (AO) algorithm is constructed to obtain a high-quality solution. Moreover, we provide a sufficient condition to ensure the feasibility of the formulated problem.
		
		\item {\color{black}After making some simplifications, we first derive the scaling order of the radar SINR in the active RIS-aided sensing systems. Subsequently, we delve into the case where the number of reflecting elements is moderate, and explore the optimal transmit power allocation between the active RIS and the radar. Additionally, we discuss the practical deployment strategies for the active RIS with respect to (w.r.t.) its position between the radar and the target. }
		
		\item Simulation results validate the potential of the active RIS in improving the performance of the ISAC system. Compared to the passive RIS, the active RIS can improve the radar SINR by up to 70 dB, since the multiplicative fading effect is greatly alleviated even when the transmit
		power of the active RIS is small. Besides, to achieve higher radar SINR, the transmit power allocation between the BS and the active RIS should be carefully chosen, and deploying the active RIS in close proximity to the intended target would be a better strategy.  
		
	\end{enumerate}
	
	The remainder of this paper is organized as follows. In Section II, we present the system model. The radar SINR maximization problem is formulated in Section III. An AO-based algorithm is developed to solve this problem in Section IV. {\color{black}In Section V, we analyze the performance of the active RIS-aided sensing system.} Finally, Sections VI, VII and VIII show the numerical results, open problems and conclusions, respectively.
	
	\emph{Notations}: Boldface lower case and upper case letters denote vectors, and matrices, respectively. ${{\mathbb{C}}^{M \times N }}$ denotes the set of $M \times N$ complex matrices. $\mathbf{X} \succeq 0$ means $\mathbf{X}$ is a positive semidefinite matrix. ${{\mathbb{E}}}{\left[\cdot\right]} $ denotes the expectation operation. ${\left\| {\mathbf{x}} \right\|_2}$ denotes the 2-norm of vector ${\mathbf{x}}$. The operation vec($\mathbf{A}$) denotes the vectorization of the matrix $\mathbf{A}$.  ${\mathbf{X}} \otimes {\mathbf{Y}}$, and ${\mathbf{X}} \odot {\mathbf{Y}}$ denote the Kronecker product and Hadamard product between $\mathbf X$ and $\mathbf Y$, respectively. ${\left\| {\mathbf{X}} \right\|_F}$ and ${\rm{Tr}}\left( {\mathbf{X}} \right)$   denote the  Frobenius norm and trace operation of ${\mathbf{X}}$, respectively. $\nabla {f_{\mathbf{x}}}\left( {\mathbf{x}} \right)$ denotes the gradient of function $f$ w.r.t. the vector ${\mathbf{x}} $. $\mathbf{0}$ denotes the zero vector and $\mathbf{I}_N$ denotes the $N \times N$ identity matrix. ${\cal C}{\cal N}({\mathbf{0}},{\mathbf{I}})$ represents a circularly symmetric complex Gaussian random vector following the distribution with zero mean and unit variance matrix. $[\mathbf{x}]_{m}$ denotes the $m$-th element of vector $\mathbf{x}$ and $[\mathbf{X}]_{p:q,m:n}$ denotes a matrix consisting of the $p$-th to the $q$-th rows and the $m$-th to the $n$-th columns of matrix $\mathbf{X}$. ${\rm{Diag}}(\mathbf{x})$ denotes a diagonal matrix that has the entries of vector $\mathbf{x}$ placed along its main diagonal, and ${\rm{diag}}\{\mathbf{X}\}$ denotes a vector composed of the main diagonal elements of matrix $\mathbf{X}$. ${\left( \cdot \right)^{\rm{T}}}$, ${\left( \cdot \right)^{*}}$, and ${\left(  \cdot \right)^{\rm{H}}}$ denote the transpose, conjugate, and Hermitian operators, respectively.

	\section{System Model}
	We consider an active RIS-aided DFRC system shown in Fig. 1, which consists of an active RIS, a radar target, communication users, and a DFRC BS. The DFRC BS is equipped with $M$ antennas, which are used to transmit communication symbols to $K$ single-antenna users and transmit radar probing waveforms to the surrounding target. The direct link between the DFRC BS and the target is assumed to be blocked{\color{black}\footnote{{\color{black}We adopt this assumption to streamline the subsequent derivation and emphasize the necessity of deploying the active RIS. When the direct link between the target and the BS is non-negligible, the received signal at the DFRC BS becomes more intricate, which contains the direct BS-Target-BS, BS-Target-RIS-BS, and BS-RIS-Target-BS links, etc. Nonetheless, the primary challenge in addressing the aforementioned problem, such as the intractable radar SINR and the optimization of quartic function, aligns with the focus of our discussion, rendering our proposed algorithm still viable.}}}, so establishing strong virtual LoS links between them is necessary. To improve the channel condition, a building's facade is equipped with an active RIS consisting of $N$ reflecting elements. For the active RIS, each reflecting element is connected to one power amplifier, and thus the active RIS can not only tune the phase shift of the incident signal but also amplify it.
	
	\begin{center}
		\begin{figure}
			\vspace{-0.3cm}
			\centering
			\includegraphics[width=3.5in]{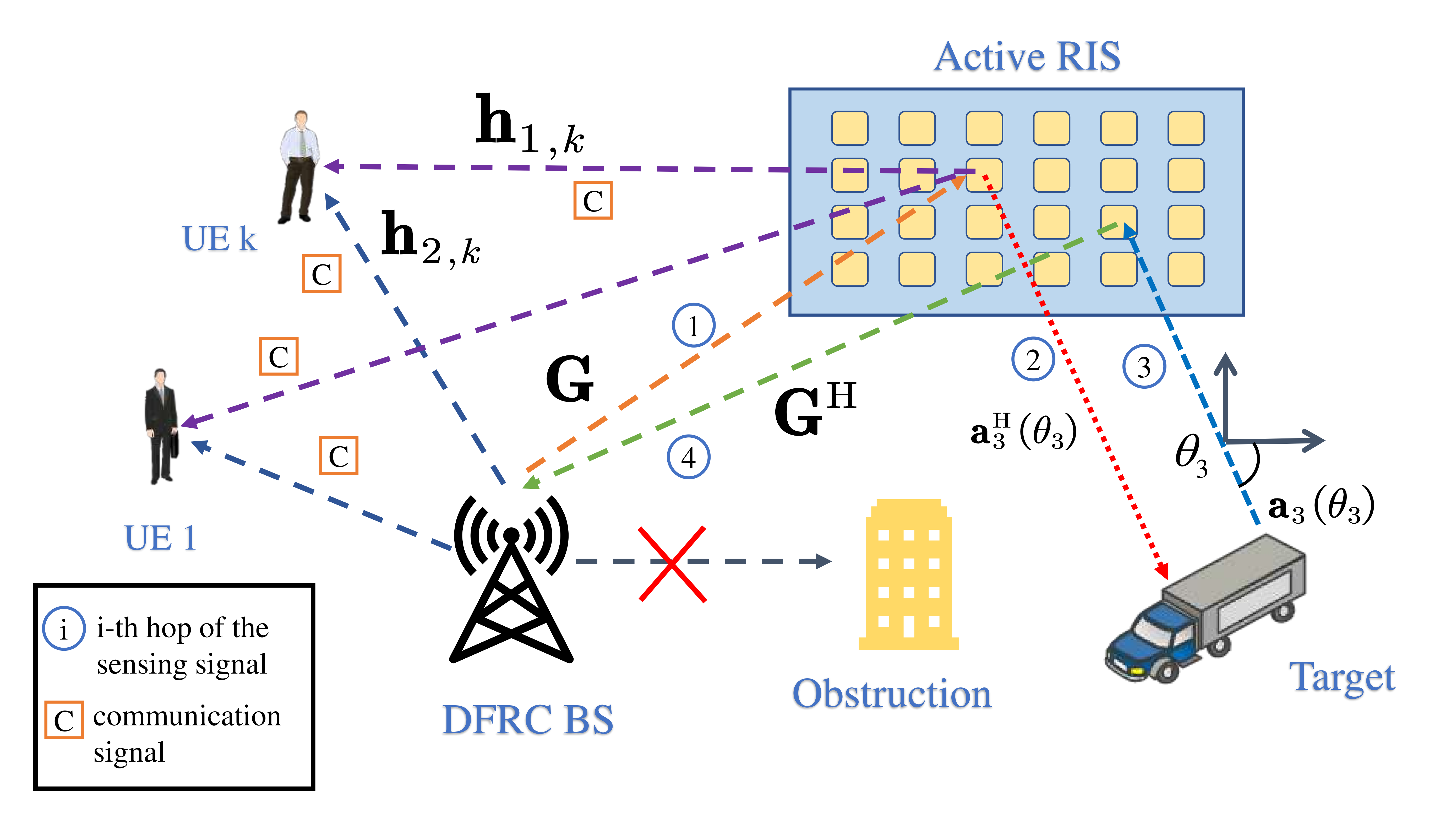}
			\caption{{\color{black}An active RIS-aided ISAC system.}}	\vspace{-0.5cm}
			\label{fig1}
		\end{figure}
		\vspace{-1cm}
	\end{center}

	\subsection{Transmission Signal}
	The transmit signal at the DFRC BS is expressed as
	\begin{equation}\label{transmit_signal}
		\begin{aligned}
			{\mathbf{x}} &= {\mathbf{W}}_{\mathrm{r}}{\mathbf{s}}+{\mathbf{W}}_{\mathrm{c}}{\mathbf{c}} \\
			&=[\begin{array}{ll}
				\mathbf{W}_{\mathrm{r}}\,, & \mathbf{W}_{\mathrm{c}}
			\end{array}][\begin{array}{ll}
				\mathbf{s}^{\mathrm{T}}, & \mathbf{c}^{\mathrm{T}}
			\end{array}]^{\mathrm{T}} 
			=\mathbf{W} \hat{\mathbf{x}},
		\end{aligned}
	\end{equation}
	where ${\mathbf{s}}\in \mathbb{C} ^{M\times1}$ denotes the radar signal and ${\mathbf{c}}\in \mathbb{C} ^{K\times1}$ denotes the transmission symbol intended to $ K $ users. Additionally,  ${\mathbf{W}}_{\mathrm{r}}=\left[{\mathbf{w}}_{{\mathrm{r}},1},{\mathbf{w}}_{{\mathrm{r}},2},\cdots,{\mathbf{w}}_{{\mathrm{r}},M} \right] \in \mathbb{C} ^{M\times M}$ and ${\mathbf{W}}_{\mathrm{c}}=\left[{\mathbf{w}}_{{\mathrm{c}},1},{\mathbf{w}}_{{\mathrm{c}},2},\cdots,{\mathbf{w}}_{{\mathrm{c}},K} \right] \in \mathbb{C} ^{M\times K} $ represent the beamforming matrices for radar and communication, respectively. {\color{black} Here ${\mathbf{W}}\in \mathbb{C} ^{M\times(M+K)}$ denotes the equivalent DFRC transmit beamforming matrix.}  {We assume that  the radar signal is generated by pseudo-random coding, which satisfies $ \mathbb{E}\left[{\mathbf{s}}\right]={\mathbf{0}} $ and $\mathbb{E}\left[ {{{\mathbf{s}}}{{{\mathbf{s}}^{\rm{H}}} }}\right] = {\mathbf{I}}_M$\cite{5}. We also assume that the transmit symbol ${\mathbf{c}}$ satisfies ${\cal C}{\cal N}\left( {{\mathbf{0}},{{\mathbf{I}}_{K}}} \right)$, and the radar and communication signals are mutually uncorrelated.} Thus, the transmit signal covariance matrix is given by
	\begin{equation} \label{signal covariance matrix }
		\begin{aligned}
			{\mathbf{R}}&={\mathbb{E}\left[\mathbf{xx}^{\rm{H}}\right]}  ={\mathbf{W}}{\mathbf{W}}^{\rm{H}} 
			={\mathbf{W}}_{\mathrm{r}}{\mathbf{W}}_{\mathrm{r}}^{\rm{H}}+\sum\limits_{k = 1}^K{{\mathbf{R}}_k} ,
		\end{aligned}
	\end{equation}
	where the rank-1 matrix ${\mathbf{R}}_k$ is introduced by ${\mathbf{R}}_k\triangleq{\mathbf{w}}_{{\mathrm{c}},k}{\mathbf{w}}_{{\mathrm{c}},k}^{\rm{H}}$.
	
	\subsection{Channel Model}
	{\color{black}Let us define $\mathbf{G}\in \mathbb{C} ^{N \times M}$, ${\mathbf{h}}_{1,k}\in \mathbb{C} ^{N \times 1}$ and ${\mathbf{h}}_{2,k}\in \mathbb{C} ^{M \times 1}$ as the channel between the DFRC BS and the active RIS, the channel between the active RIS and the $k$-th user (UE), and the channel between the DFRC BS and the $k$-th UE, respectively. We assume that the channel state information (CSI) of the above channels is perfectly known at the DFRC BS by applying the effective channel estimation methods\cite{52}. 
		
		We consider a general urban micro-cellular (UMi) scenario, and the large-scale path loss for communication links $\beta_\mathrm{c}$ can be modeled as
		\begin{equation} \label{sthbrr}
			\beta_\mathrm{c}{[\mathrm{dB}]}=  10 \alpha \log _{10}\left(\frac{d_s}{d_0}\right)+\beta 
			+10 \gamma \log _{10}\left(\frac{f}{f_0}\right)+\chi_{\sigma}^{\mathrm{ABG}},
		\end{equation}
		where $\alpha$ and $\gamma$ are coefficients showing the dependence of path loss on distance and frequency, respectively, and $\beta$ represents an optimized offset value for path loss in decibels. The symbol $d_s$ denotes the  transmitter-receiver separation distance in meters, while $f$ indicates the carrier frequency in GHz. Here, $d_0=1$ m and $f_0=1 $ GHz are the reference distance and reference frequency, respectively. The shadow fading (SF) standard deviation, $\chi_{\sigma}^{\mathrm{ABG}}$, describes the large-scale signal fluctuations about the mean path loss over distance. A comprehensive list of these parameters can be found in\cite[Table 1]{channel_model}.  
		
		Since the communication users are typically located on the ground with densely surrounded scatters, Rayleigh channel models are adopted for both ${\mathbf{h}}_{1,k}$ and ${\mathbf{h}}_{2,k}$. The Rician channel model is used for the BS-RIS channel {\color{black}${\mathbf{G}}$}, which encompasses both the non-LoS (NLoS) Rayleigh fading component ${\mathbf{G}}_{\rm{NLoS}}$ and the LoS component ${\mathbf{G}}_{\rm{LoS}}$.
		The NLoS fading component ${\mathbf{G}}_{\rm{NLoS}}$ can be expressed as $\mathbf{G}_{\rm{NLoS}}\sim \mathcal{C N}(\mathbf{0}, \boldsymbol{\Sigma}_{\mathrm{R}} \otimes \boldsymbol{\Sigma}_{\mathrm{B}})$, where  $\boldsymbol{\Sigma}_{\mathrm{B}}\succeq 0$ and $\boldsymbol{\Sigma}_{\mathrm{R}}\succeq 0$ denote the spatial correlation matrices with unit diagonal elements at the DFRC BS and the active RIS for channel ${\mathbf{G}}_{\rm{NLoS}}$, respectively. The LoS component ${\mathbf{G}}_{\rm{LoS}}$ is determined by the DoAs of the active RIS $\theta_2$ and the angles-of-departure (AoDs) of the BS $\theta_1$, which is given by
		\begin{equation}\label{chanel los}
			{\mathbf{G}}_{\rm{LoS}}={\mathbf{a}}_2\left( \theta _2 \right) {\mathbf{a}}_{1}^{\rm{H}}\left( \theta _1 \right) .
		\end{equation} 
		Vectors ${\mathbf{a}}_1\left( \theta_1\right) $ and ${\mathbf{a}}_2\left( \theta_2\right) $ are respectively the array response vectors of the BS antennas as well as the RIS, which are mathematically formulated as
		\begin{equation}\label{steering_vector}
			\begin{aligned}
				\mathbf {a}_1(\theta _1)&=[1,e^{-j2\pi d_{\rm{BS}}  \frac{\sin (\theta _1)}{\lambda }},{\ldots },e^{-j2\pi d_{\rm{BS}} (M-1) \frac{\sin (\theta _1)}{\lambda }}]^{\rm{T}} ,\\
				\mathbf {a}_2(\theta _2)&=[1,e^{-j2\pi d_{\rm{RIS}}  \frac{\sin (\theta _2)}{\lambda }},{\ldots },e^{-j2\pi d_{\rm{RIS}} (N-1) \frac{\sin (\theta _2)}{\lambda }}]^{\rm{T}} ,
			\end{aligned}
		\end{equation}
		where $\lambda$ represents the wavelength of the carrier wave, $d_{\rm{BS}}$ and $d_{\rm{RIS}}$ denote the antenna spacing distance of the BS and the spacing between the adjacent reflecting elements of the active RIS, respectively. For simplicity, we set $d_{\rm{BS}}=d_{\rm{RIS}}=\frac{\lambda }{2}$.
		
		Assuming that the spatial extent of the target is relatively small, the incident signal is reflected only by the point target. Thus, the target response matrix between the RIS and the target can be denoted by ${\mathbf{A}}={\beta_{\mathrm{r}}}\mathbf{a}_3\left( \theta _3 \right) \mathbf{a}_{3}^{\rm{H}}\left( \theta _3 \right) \in \mathbb{C} ^{N \times N}$, where $ \theta _3$ is the DoA w.r.t. the active RIS, and the steering vector $\mathbf{a}_3\left( \theta _3 \right)$ is defined similarly as those in (\ref{steering_vector}). The amplitude of the complex coefficient $\tilde{\beta_{\mathrm{r}}}$, characterizing the multiplicative fading of the RIS-target-RIS channel, is modeled using the radar range equation for evaluation. If the target can be regarded as a single scatter object, and various additional loss and gain factors are ignored for simplicity, the received power $P_{{r}}$ at the radar can be calculated using the following equation:
		\begin{equation}\label{hre}
			P_{r}=\frac{P_{t} G^{2} \lambda^{2} S}{(4 \pi)^{3} R^{4}},
		\end{equation} 
		where $P_{t}$ and $G$ represent the transmit power and the antenna gain of radar, respectively. Parameter $S$ represents the radar cross section (RCS) of the target, and $R$ denotes the distance between the radar and the target.  Since the active RIS can be regarded as a monostatic multiple-input and multiple-output (MIMO) radar,  the path loss of the RIS-target-RIS link $\beta_{\mathrm{r}}$ can be modeled as
		\begin{equation}\label{grgh}
			\beta_{\mathrm{r}}=|\tilde{\beta_{\mathrm{r}}}|=\sqrt{\frac{ \lambda^{2} S}{(4 \pi)^{3} R^{4}}}. 
		\end{equation}
	}
	{\vspace{-0.5cm}}

	\subsection{Active RIS Model}
	Different from the passive RIS, which only comprises a multitude of passive elements, the active RIS is outfitted with active reflection-type amplifiers on each of its constituent elements. Thus, the active RIS can further amplify the reflected signal, resulting in an enhanced performance for sensing and communication tasks. {The reflecting coefficient matrix of the active RIS is denoted by $\mathbf{\Phi} ={\rm{Diag}}([ {v_1, \cdots ,{v_n}}, \cdots ,{v_N}])$, and the power amplification gain $|v_n|^2,n=1,2,\cdots,N$ should be less than the maximum power amplification gain $a_{\rm{RIS}}$.} However, owing to the adoption of active components, the thermal noise and the power consumption of the active RIS are non-negligible compared to those of the passive RIS.
	In the considered scenario, the RIS first reflects the transmit signal from the BS to communication users and the target. Then, the echo signal reflected by the target is reflected by the RIS to the BS. {\color{black}By neglecting the static noise like in [29],} the first and second reflected signals can be respectively  expressed as
	\begin{align}
		{\mathbf{y}}_{1}^{\rm{r}}&={\mathbf{\Phi Gx}}+{\mathbf{\Phi v_1}},  \label{first reflection}\\
		{\mathbf{y}}_{2}^{\rm{r}}&={\mathbf{\Phi }} ^{\rm{H}}{\mathbf{A\Phi Gx}}+{\mathbf{\Phi }} ^{\rm{H}}{\mathbf{A\Phi v}}_1+{\mathbf{\Phi }}  ^{\rm{H}}{\mathbf{v}} _2.\label{vbf}
	\end{align}
	Vectors ${\mathbf{v}}_1$ and ${\mathbf{v}}_2 $ denote the additive white Gaussian noise (AWGN) at the active RIS, which follow the same distribution of ${\cal C}{\cal N}\left( {{\mathbf{0}},\sigma^2{{\mathbf{I}}_{N}}} \right)$ with the noise power of $\sigma^2$, respectively. Denote $P_{\rm{RIS}}$ by the maximum RIS transmit power, the transmit power constraint of the active RIS is given by	
	\begin{equation}\label{RIS constrains}
		\mathbb{E}\left[{\left\|{\mathbf{y}}_{1}^{\rm{r}}\right\|}^{2}_2+\left\|{\mathbf{y}}_{2}^{\rm{r}}\right\|^{2}_2\right] \le P_{\rm{RIS}}.
	\end{equation}	
	By substituting (\ref{first reflection}) and (\ref{vbf}) into (\ref{RIS constrains}), we have
	\begin{align} \label{RIS constraint}
		{\operatorname{Tr}}&(\mathbf{\Phi^{\rm{H}} A \Phi G R G^{\rm{H}} \Phi^{\rm{H}} A^{\rm{H}} \Phi})+\sigma^2{\operatorname{Tr}}(\mathbf{\Phi^{\rm{H}} A \Phi \Phi^{\rm{H}} A^{\rm{H}} \Phi}) \notag \\
		&+{\operatorname{Tr}}(\mathbf{\Phi G R G^{\rm{H}} \Phi^{\rm{H}}} )+2\sigma^2{\operatorname{Tr}}(\mathbf{\Phi \Phi ^{\rm{H}}} )\le P_{\rm{RIS}}.
	\end{align}

	\section{Problem Formulation}
	
	\subsection{Radar Performance Metric}
	{\color{black} In sensing tasks, the received echo signals at the DFRC BS encompass the thermal noise originating from both the active RIS and the DFRC BS, along with the signal transmitted through the BS-RIS-BS link and the BS-RIS-Target-RIS-BS link. Given that the echo signal from the BS-RIS-BS link lacks any pertinent information about the target, it can be considered as interference with the target sensing. Thus, the received radar signal at the DFRC BS can be expressed as 
		\begin{equation}\label{received echo}
			\begin{aligned}
				\tilde{{\mathbf{y}}}_{\rm{r}}&={\mathbf{G}}^{\rm{H}}({\mathbf{y}}_{1}^{\rm{r}}+{\mathbf{y}}_{2}^{\rm{r}})+{\mathbf{z}}_{\rm{r}}\\
				&={\mathbf{G}}^{\rm{H}}{\mathbf{\Phi }} ^{\rm{H}}{\mathbf{A}}{\mathbf{\Phi }} {\mathbf{G}}{\mathbf{x}}+{\mathbf{G}}^{\rm{H}}{\mathbf{\Phi }} ^{\rm{H}}{\mathbf{A}}{\mathbf{\Phi }} {\mathbf{v}}_1 
				+ {\mathbf{G}}^{\rm{H}}{\mathbf{\Phi }}{\mathbf{v}}_1\\
				&\quad+{\mathbf{G}}^{\rm{H}}{\mathbf{\Phi }} ^{\rm{H}}{\mathbf{v}}_2+\underbrace{{ \mathbf{G}}^{\rm{H}}\mathbf{\Phi}\mathbf{ G} {\mathbf{x}}}_{\textrm{Interference echo}}+{\mathbf{z}}_{\rm{r}},
			\end{aligned}
		\end{equation}
		where ${\mathbf{z}}_{\rm{r}}$ denotes the AWGN at the DFRC BS, which follows the distribution of ${\cal C}{\cal N}\left( {{\mathbf{0}},\sigma_{\rm{r}}^2{{\mathbf{I}}_{M}}} \right)$  with the noise power of $\sigma_{\rm{r}}^2$. 
		Since the interference term $\mathbf{G}^{\rm{H}}\mathbf{\Phi}\mathbf{ G} {\mathbf{x}}$ is known at the  DFRC BS, after adopting effective self-interference (SI) cancellation techniques, 
		the received radar signal is given by
		\begin{equation}\label{received echo2}
			\begin{aligned}
				{{\mathbf{y}}_{\rm{r}}}
				&={\mathbf{G}}^{\rm{H}}{\mathbf{\Phi }} ^{\rm{H}}{\mathbf{A}}{\mathbf{\Phi }} {\mathbf{G}}{\mathbf{x}}+{\mathbf{G}}^{\rm{H}}{\mathbf{\Phi }} ^{\rm{H}}{\mathbf{A}}{\mathbf{\Phi }} {\mathbf{v}}_1 
				+ {\mathbf{G}}^{\rm{H}}{\mathbf{\Phi }}{\mathbf{v}}_1\\
				&\quad +{\mathbf{G}}^{\rm{H}}{\mathbf{\Phi }} ^{\rm{H}}{\mathbf{v}}_2+{{\eta \mathbf{G}}^{\rm{H}}\mathbf{\Phi}\mathbf{ G} {\mathbf{x}}}+{\mathbf{z}}_{\rm{r}},
			\end{aligned}
		\end{equation}
		where $\eta$ denotes the SI coefficient after mitigation.   
		Defining ${\mathbf{B}}\triangleq{\mathbf{G}}^{\rm{H}}{\mathbf{\Phi }} ^{\rm{H}}{\mathbf{A\Phi G }} \in \mathbb{C} ^{M \times M}$ and ${\mathbf{C}}\triangleq{\mathbf{G}}^{\rm{H}}{\mathbf{\Phi }} ^{\rm{H}}{\mathbf{A\Phi}}\in \mathbb{C} ^{M \times N} $, the radar SINR can be formulated as\cite{SNR_Reason,10}{\color{black}\footnote{{\color{black}The SINR is an essential performance metric of radar sensing\cite{19},  intimately linked with the pairwise Kullback-Leibler (KL) divergences between the densities of observations under the two alternative hypotheses. In light of the potential degradation of sensing performance resulting from thermal noise and interference from the active RIS, we use the radar SINR as a performance metric. Besides, the beampattern is evaluated in the simulation section to further assess the radar sensing performance.}}}
		\begin{equation}\label{SINR}
			{\rm{ SINR}}={\operatorname{Tr}}(\mathbf{BRB}^{\rm{H}}\mathbf{J} ^{-1}). 
		\end{equation} 
		The interference-plus-noise covariance matrix ${\mathbf{J}}$ is given by
		\begin{equation}\label{noise covariance matrix}
			\begin{aligned}
				\mathbf{J}&=\mathbf{D}+\mathbf{ERE}^{\rm{H}},
			\end{aligned}
		\end{equation}
		where $ \mathbf{D}\triangleq\sigma^2({\mathbf{G}}^{\rm{H}}{\mathbf{\Phi }}{\mathbf{\Phi }} ^{\rm{H}}{\mathbf{A}}{\mathbf{\Phi G}} +\mathbf{G}^{\rm{H}}{\mathbf{ \Phi }} ^{\rm{H}}{\mathbf{A \Phi }}{\mathbf{\Phi }}^{\rm{H}}\mathbf{G })+ \sigma^2\mathbf{CC} ^{\rm{H}} +2\sigma^2\mathbf{G} ^{\rm{H}}\mathbf{\Phi}^{\rm{H}}\mathbf{\Phi G} +\sigma_{\rm{r}}^2\mathbf{I} _M$  is the equivalent noise covariance matrix and $ \mathbf{E}=\eta \mathbf{G} ^{\rm{H}} \mathbf{\Phi G} $.}

	\subsection{Communication Performance Metric}
	For communication tasks, the signal transmitted from the BS to the user passes through two paths: the BS-UE direct link and the BS-RIS-UE link. As a result, the signal received by the $k$-th user can be represented as: 
	\begin{equation}\label{UE_k}
		\begin{aligned}
			y_{{\mathrm{c}},k}={\mathbf{h}}_{1,k}^{\rm{H}}{\mathbf{\Phi Gx}} +{\mathbf{h}}_{1,k}^{\rm{H}}{\mathbf{\Phi v}} _1 +{\mathbf{h}}_{2,k}^{\rm{H}}{\mathbf{x}}+z_k,   
		\end{aligned}
	\end{equation}
	where $z_k$ is the AWGN with variance of $\sigma _{\rm{z}}^{2}$. The SINR of the $ k $-th user can be expressed as   
	\begin{equation}\label{SINR_k}
		{\rm{SINR}}_k=\frac{{\mathbf{h}}_{k}^{\rm{H}}{\mathbf{R}}_k{\mathbf{h}}_k}{{\mathbf{h}}_{k}^{\rm{H}}\left( {\mathbf{R}}-{\mathbf{R}}_k \right) {\mathbf{h}}_k+\sigma^{2}{\mathbf{h}}_{1,k}^{\rm{H}}{\mathbf{\Phi} }{\mathbf{\Phi} }^{\rm{H}}{\mathbf{h}}_{1,k}  +\sigma _{\rm{z}}^{2}},
	\end{equation}
	where ${\mathbf{h}}_k^{\rm{H}}\triangleq{\mathbf{h}}_{1,k}^{\rm{H}}{\mathbf{\Phi G}} +{\mathbf{h}}_{2,k}^{\rm{H}}$ can be regarded as the equivalent channel between the BS and the $k$-th user.

	\subsection{Optimization Problem}
	In this work, we aim to maximize the received radar SINR at the DFRC BS by jointly optimizing the beamforming matrix $ {\mathbf{W}} $ at the BS and the reflecting coefficient matrix ${\mathbf{\Phi}} $ at the RIS. Accordingly, the problem is formulated as
	\begin{subequations}\label{optimize}
		\begin{align}
			\mathop {\max }\limits_{{{\mathbf{W}},{\mathbf{\Phi}} } } \quad  & {\rm{SINR}} \label{bfg} 
			\\
			\textrm{s.t.}\qquad
			&{\rm{SINR}}_k  \ge {\mathbf{\xi}},k = 1, \cdots ,K,\label{communication constrain}\\
			&{\operatorname{Tr}}\left( {\mathbf{R}} \right) \le P_{\rm{BS}}, \label{BS power1}\\
			&\mathbb{E}\left[{\left\|{\mathbf{y}}_{1}^{\rm{r}}\right\|}_2^{2}+\left\|{\mathbf{y}}_{2}^{\rm{r}}\right\|_2^{2}\right] \le P_{\rm{RIS}},\label{RIS power} \\
			&|v_n|^2\le a_{\rm{RIS}} , n = 1, \cdots ,N,	
		\end{align}
	\end{subequations}
	where $P_{\rm{BS}}$ denotes the maximum transmit power and $\xi$ is the required SINR for all communication users.
	
	The resolution of Problem (\ref{optimize}) poses several challenges, primarily due to the presence of the quartic objective function (\ref{SINR}) and quartic constraint (\ref{RIS power}). Additionally, the coupled optimization variables and the nonconvex constraint (\ref{communication constrain}) further aggravate the challenge.
	
	\section{Proposed Solver via MM algorithm}
	This section aims to address the aforementioned difficulties by initially reformulating the objective function into a more tractable form. Subsequently, we employ the AO method to solve the reformulated problem. In particular, the optimization of beamforming matrices is accomplished by using an MM-semidefinite programming (SDP) algorithm, whereas the reflecting coefficients are optimized by reformulating the problem as a sum of square form (SOS) expression and then solving it through the SDR algorithm.
	
	\subsection{Reformulation of the Objective Function}
	We employ the MM algorithm\cite{14} as a means of resolving Problem (\ref{optimize}). The main idea of the MM algorithm is to devise a sequence of tractable approximate subproblems to tackle the non-convex problem. Specifically, assuming that $f(\mathbf{t})$ is the original objective function that needs to be maximized, the surrogate function $\tilde{f}(\mathbf{t}|\mathbf{t}_i)$ to lower bound the objective function at the $i$-th iteration should satisfy the following conditions:
	\begin{enumerate}
		\item $\tilde{f}(\mathbf{t}_i|\mathbf{t}_i)=f(\mathbf{t}_i)$; \label{ndd}
		\item$\nabla_{\mathbf{t}}\tilde{f}(\mathbf{t}|\mathbf{t}_i)|_{{\mathbf{t}}={\mathbf{t}}_i}=\nabla_{\mathbf{t}}f(\mathbf{t}) |_{{\mathbf{t}}={\mathbf{t}}_i}$; \label{mud}
		\item $\tilde{f}(\mathbf{t}|\mathbf{t}_i) \le f(\mathbf{t})$. \label{mfm}
	\end{enumerate}
	Based on the MM framework, we obtain the lower bound of the objective function by the first-order Taylor expansion, which is shown in Lemma 1.
	
	\itshape \textbf{Lemma 1:} \upshape The radar SINR in (\ref{SINR}) can be minorized by the surrogate function at the $i$-th iteration given by
	\begin{equation} \label{awsl}
		\begin{aligned}
			\operatorname{Tr}\left(\mathbf{X}^{\rm{H}} \mathbf{J}^{-1}\mathbf{X}\right) 
			&\geq 2\operatorname{Re}(\operatorname{Tr}\left(\mathbf{X}^{\rm{H}}_i\mathbf{J}^{-1}_i\mathbf{X}\right)) \\
			&\quad -\operatorname{Tr}\left(\mathbf{J}^{-1}_i \mathbf{X}_i \mathbf{X}^{\rm{H}}_i\mathbf{J}^{-1}_i\mathbf{J}\right),
		\end{aligned}
	\end{equation}
	where $ \mathbf{X}_i=\mathbf{B}_i\mathbf{W}_i $ denotes the auxiluary matrix $ \mathbf{X}=\mathbf{BW} $ at the $ i $-th iteration, and $ \mathbf{J}_i $ denotes the interference-plus-noise covariance matrix $ \mathbf{J} $ at the $ i $-th iteration, respectively. 
	
	\itshape \textbf{Proof:}  \upshape Please refer to Appendix A. $\hfill\blacksquare$
	
	Next, we will introduce an iterative methodology that uses Lemma 1 to solve the problem. Given the strong correlation between the optimization variables $\mathbf{W}$  and $\mathbf{\Phi}$, the AO algorithm has been employed in the forthcoming research.
	
	\subsection{Optimize $\mathbf{W}$}
	Within this subsection, we undertake the optimization of the beamforming matrix ${\mathbf{W}}$, keeping the reflecting coefficient matrix ${\mathbf{\Phi}}$  constant. By using Lemma 1, the surrogate objective function at the $i$-th iteration is
	\begin{equation} \label{gtah}
		\begin{aligned}
		f(\mathbf{W}|\mathbf{W}_i)&=2\operatorname{Re}(\operatorname{Tr}\left(\mathbf{W}^{\rm{H}}_i\mathbf{B}^{\rm{H}}\mathbf{J}^{-1}_i\mathbf{BW}\right)) \\
			&\quad -\operatorname{Tr}\left(\mathbf{J}^{-1}_i \mathbf{BR}_i \mathbf{B}^{\rm{H}}\mathbf{J}^{-1}_i\mathbf{J}\right).
		\end{aligned}
	\end{equation}
	Next, the transmit power constraint of the active RIS in (\ref{RIS power}) can be rewritten as  
	\begin{equation}\label{sdfwr}
		\left\|  {{{{\mathbf{\Phi}}}^{\rm{H}}}\mathbf{A\Phi G}\mathbf{W}  }  \right\|^2_F+\left\|  \mathbf{\Phi G}\mathbf{W}    \right\|^2_F\le e,
	\end{equation}
	where the constant $e$ is given by $ e \triangleq P_{\rm{RIS}}-2\sigma^2{\operatorname{Tr}}(\mathbf{\Phi \Phi ^{\rm{H}}} )-\sigma^2{\operatorname{Tr}}(\mathbf{\Phi^{\rm{H}} A \Phi \Phi^{\rm{H}} A^{\rm{H}} \Phi}) $.
	
	The QoS constraints of the communication UEs in (\ref{communication constrain}) can be equivalently transformed as
	\begin{equation} 
		\begin{aligned}
			(1+{\mathbf{\xi}}^{-1}){\mathbf{h}}_k^{\rm{H}}{\mathbf{w}}_{M+k}{\mathbf{w}}_{M+k}^{\rm{H}}{\mathbf{h}}_k& \ge {\mathbf{h}}_k^{\rm{H}}{\mathbf{R}}{\mathbf{h}}_k+d_k, \quad  \\&k=1,2,\cdots,K, \label{communication constrain4}
		\end{aligned}
	\end{equation}
	where $d_k\triangleq\sigma^{2}{\mathbf{h}}_{1,k}^{\rm{H}}{\mathbf{\Phi} }{\mathbf{\Phi} }^{\rm{H}}{\mathbf{h}}_{1,k}+\sigma _{\rm{z}}^{2}$ is a constant independent of $\mathbf{W}$.
	The nonconvex constraint (\ref{communication constrain4}) is a difference-of-convex (DC) programming problem that can be solved via the successive convex approximation (SCA) effectively.
	Combining with (\ref{gtah}) and (\ref{sdfwr}), the equivalent subproblem corresponding to the beamforming matrix $\mathbf{W}$ is rewritten as
	\begin{subequations}\label{optimize for precoding2}
		\begin{align}
			\mathop {\max }\limits_{{{\mathbf{W}} } }  \quad &f(\mathbf{W}|\mathbf{W}_i) \label{absd}
			\\
			\ \textrm{s.t.} \quad
			&(1+{\mathbf{\xi}}^{-1})(2\operatorname{Re}(\mathbf{h} _k^{\rm{H}}{\mathbf{w}}_{M+k,i}{\mathbf{w}}_{M+k}^{\rm{H}}\mathbf{h}_k) \nonumber \\ 
			&-\mathbf{h} _k^{\rm{H}}{\mathbf{w}}_{M+k,i}{\mathbf{w}}_{M+k,i}^{\rm{H}}\mathbf{h} _k) \nonumber \\ 
			&\ge {\mathbf{h}}_k^{\rm{H}}{\mathbf{R}}{\mathbf{h}}_k+d_k, \quad 
			k=1,2,\cdots,K,\label{communication constrain3}\\
			&(\ref{sdfwr}), \, (\ref{BS power1}).
		\end{align}
	\end{subequations}
	The convexity of the aforementioned problem can be verified, thereby rendering it an SDP problem amenable to resolution through CVX tools \cite{grant2014cvx}. 
	
	The convergence of the MM algorithm is affected by the initial point, so we propose a scheme to find an initial point that is a feasible point of Problem (\ref{optimize for precoding2}). Inspired by \cite{5}, we first construct a feasibility problem corresponding to $\mathbf{R}_k$ and $\mathbf{R}$, then the feasible initial point $\mathbf{W}_0$ can be obtained by decomposing  $\mathbf{R}_k$ and $\mathbf{R}$. In particular, the feasible problem is formulated as follows
	\begin{subequations}\label{optimize for precoding7}
		\begin{align}
			\textrm{find }\quad &{\mathbf{R}}, {\mathbf{R}}_{k, \, k=1,...,K}
			\\
			\quad \textrm{s.t.}  \; \quad
			&(\ref{BS power1}),\;(\ref{sdfwr})\\
			&(1+{\mathbf{\xi}}^{-1}){\mathbf{h}}_k^{\rm{H}}{\mathbf{R}}_k{\mathbf{h}}_k \ge {\mathbf{h}}_k^{\rm{H}}{\mathbf{R}}{\mathbf{h}}_k+d_{k}, \label{communication constrain7}\\
			&\mathrm{Rank}({\mathbf{R}}_k)=1,\quad \mathbf{R} \succeq 0, \quad \mathbf{R}_{k}\succeq 0,\quad \label{gra}\\
			&\mathbf{R}-\sum_{k=1}^{K} \mathbf{R}_{k} \succeq 0, \quad k=1,...,K. \label{bsndnhhs}
		\end{align}
	\end{subequations}
	By dropping the rank-one constraints (\ref{gra}), the relaxed problem is convex and can be solved via CVX. Denoting $\tilde{\mathbf{R}}$ and  $ \tilde{\mathbf{R}}_{k} $ as a feasible solution of Problem (\ref{optimize for precoding7}), we can construct a beamforming matrix $\hat{\mathbf{W}}_{{\mathrm{c}}}=\left[\hat{\mathbf{w}}_{{\mathrm{c}},1},\hat{\mathbf{w}}_{{\mathrm{c}},2},\cdots,\hat{\mathbf{w}}_{{\mathrm{c}},K} \right] $ as follows:
	\begin{equation}\label{wck}
		\begin{aligned}
			\hat{\mathbf{w}}_{{\mathrm{c}}{,k}}=\frac{\tilde{\mathbf{R}}_{k} \mathbf{h}_{k}}{\sqrt{\mathbf{h}_{k}^{\mathrm{H}} \tilde{\mathbf{R}}_{k} \mathbf{h}_{k}}}, 
			\, \hat{\mathbf{R}}_{k}=\hat{\mathbf{w}}_{{\rm{c}}{,k}} \hat{\mathbf{w}}_{{\mathrm{c}},k}^{\mathrm{H}}, \, k=1, \cdots, K,
		\end{aligned}
	\end{equation} 
	and a beamforming matrix ${\hat{\mathbf{W}}}_{\mathrm{r}}=\left[\hat{\mathbf{w}}_{{\rm{r}},1},\hat{\mathbf{w}}_{{\rm{r}},2},\cdots,\hat{\mathbf{w}}_{{\rm{r}},N} \right] $ using the Cholesky decomposition as
	\begin{equation}\label{wr}
		\hat{\mathbf{W}}_{\mathrm{r}}\hat{\mathbf{W}}_{\mathrm{r}}^{\mathrm{H}}=\tilde{\mathbf{R}}-\sum_{k=1}^{K} \hat{\mathbf{R}}_{k}.
	\end{equation}
	In the following, we prove that (\ref{wck}) and (\ref{wr}) are feasible solutions for Problem (\ref{optimize for precoding7}). It can be seen that constraints (24b) and (24d) hold obviously with the solutions, then we only need to check constraints (\ref{communication constrain7}) and (\ref{bsndnhhs}).
	
	Constraint (\ref{communication constrain7}) holds with $  \hat{\mathbf{R}}_{k} $ since $ \mathbf{h}_{k}^{\mathrm{H}} \hat{\mathbf{R}}_{k} \mathbf{h}_{k}=\mathbf{h}_{k}^{\mathrm{H}} \tilde{\mathbf{R}}_{k} \mathbf{h}_{k} $. 
	For constraint (\ref{bsndnhhs}), according to the Cauchy-Schwarz inequality, for any $\mathbf{u} \in \mathbb{C} ^{M \times 1}$, we have
	\begin{equation}
		\begin{aligned}
			\left(\mathbf{h}_{k}^{\rm{H}} \tilde{\mathbf{R}}_{k} \mathbf{h}_{k}\right)\left(\mathbf{u}^{\rm{H}} \tilde{\mathbf{R}}_{k} \mathbf{u}\right)&=\left\|{\tilde{\mathbf{R}}_{k}}^{\frac{1}{2}} \mathbf{h}_{k}\right\|_{2}^{2}\left\|\mathbf{u}^{\mathrm{H}} {\tilde{\mathbf{R}}_{k}}^{\frac{1}{2}}\right\|_{2}^{2} \\
			&\geq\left|\mathbf{u}^{\rm{H}} \tilde{\mathbf{R}}_{k} \mathbf{h}_{k}\right|^{2},
		\end{aligned}
	\end{equation}
	which further leads to
	\begin{equation}
		\begin{aligned}
			\mathbf{u}^{\rm{H}}\left(\tilde{\mathbf{R}}_{k}-\hat{\mathbf{R}}_{k}\right) \mathbf{u}&=\mathbf{u}^{\rm{H}} \tilde{\mathbf{R}}_{k} \mathbf{u}-\left(\mathbf{h}_{k}^{\rm{H}} \tilde{\mathbf{R}}_{k} \mathbf{h}_{k}\right)^{-1}\left|\mathbf{u}^{\rm{H}} \tilde{\mathbf{R}}_{k} \mathbf{h}_{k}\right|^{2} \\
			&\ge 0. 
		\end{aligned}
	\end{equation}
	Since $ \hat{\mathbf{R}}-\sum_{k=1}^{K} \tilde{\mathbf{R}}_{k} \succeq 0 $, we have 
	\begin{equation}
		\hat{\mathbf{R}}-\sum_{k=1}^{K} \hat{\mathbf{R}}_{k}=\tilde{\mathbf{R}}-\sum_{k=1}^{K} \tilde{\mathbf{R}}_{k}+\sum_{k=1}^{K}\left(\tilde{\mathbf{R}}_{k}-\hat{\mathbf{R}}_{k}\right) \succeq 0.
	\end{equation}
	
	Therefore, it is demonstrated that the constructed solution 	$\hat{\mathbf{w}}_{{\mathrm{c}},k}$ and $	\hat{\mathbf{W}}_{\mathrm{r}}$ are feasible solution to Problem (\ref{optimize for precoding7}). Besides, to achieve a better converged solution, we propose a transformation of Problem (\ref{optimize for precoding7}) by replacing the QoS constraints (\ref{communication constrain7}) with a tighter constraint. Specifically, we use the constraint 	${\mathbf{h}}_k^{\rm{H}}(1+{\mathbf{\xi}}_2^{-1}){\mathbf{R}}_k{\mathbf{h}}_k \ge {\mathbf{h}}_k^{\rm{H}}{\mathbf{R}}{\mathbf{h}}_k+d_k$ instead of (\ref{communication constrain7}), where ${\mathbf{\xi}}_2$ is larger than ${\mathbf{\xi}}$. {\color{black}Under tighter QoS constraints, the BS is encouraged to prioritize directing its energy towards the communication users during the optimization of BS beamforming matrices. Subsequently, as we proceed to optimize the reflecting coefficient matrix of the active RIS, it becomes easier to meet the communication QoS requirements. This, in turn, enables the active RIS to effectively allocate the energy it receives towards the sensing target. Hence, this transformation is expected to lead to a better solution for Problem (\ref{optimize for precoding7}).} 
	
	{In the following, we provide a sufficient condition that Problem (\ref{optimize for precoding7}) is feasible by considering a simplified system design. In the simplified system design, the active RIS solely amplifies the received signal while preserving its original direction. This is achieved by setting the reflecting coefficient matrix of the RIS as $\mathbf{\Phi}=\rho \mathbf{I}_N$, where $\rho$ is a real value. Based on the above settings, the term $d_k$ is reformulated as $\tilde{{d}}_k={{\mathbf{h}}_{1,k}^{\rm{H}}{\mathbf{h}}_{1,k}\rho ^2\sigma^{2}+\sigma _{\rm{z}}^{2}}$, and the equivalent communication channel for the $k$-th UE $\tilde{\mathbf{h}}_k$ is expressed as $\tilde{\mathbf{h}}_k={\mathbf{h}}_{2,k}+\rho {\mathbf{Gh}}_{1,k}$. We also construct the equivalent communication channel matrix $\tilde{\mathbf{H}}$ as $\tilde{\mathbf{H}}=[\tilde{\mathbf{h}}_1,\cdots,\tilde{\mathbf{h}}_K]$.}
	
	\itshape \textbf{Lemma 2:} \upshape Problem (\ref{optimize for precoding7}) is feasible if there exists $\rho$ that satisfies the following conditions: 
	\begin{enumerate}
		\item $\operatorname{Rank}(\tilde{\mathbf{H}})=K$,
		\item $\operatorname{Tr}(\operatorname{Diag}{([\tilde{{d}}_1,\cdots,\tilde{{d}}_K])}(\tilde{\mathbf{H}}^{\rm{H}}\tilde{\mathbf{H}})^{-1}) \le \frac{P_{\rm{BS}}}{\xi}$, \label{nd}	
		\item  $1 \le \rho  \le \sqrt{a_{\rm{RIS}}}$,
		\item $\rho^4\|\mathbf{AG}\mathbf{W}^{\star}\|_F^2+\rho^2\|\mathbf{G}\mathbf{W}^{\star}\|_F^2+\rho^4\sigma^2\operatorname{Tr}(\mathbf{AA}^{\rm{H}})+2\rho^2\sigma^2 \le P_{\rm{RIS}}$, 		
	\end{enumerate} 
	where 
	\begin{equation}
		\mathbf{W}^{\star}=\sqrt{\xi}\operatorname{Diag}{([\tilde{{d}}_1,\cdots,\tilde{{d}}_K])}^{\frac{1}{2}}\;\tilde{\mathbf{H}}(\tilde{\mathbf{H}}^{\rm{H}}\tilde{\mathbf{H}})^{-1}.
	\end{equation}
	
	\itshape \textbf{Proof:} \upshape Please refer to Appendix B. $\hfill\blacksquare$

	\subsection{{Optimize $\mathbf{\Phi}$}}
	
	In this subsection, we optimize the reflecting coefficient matrix $\mathbf{\Phi}$ when the beamforming matrix $\mathbf{W}$ is fixed. It is observed that both the transmit power of the active RIS (\ref{RIS power}) and the radar SINR (\ref{bfg}) contain the quartic function of the reflecting coefficients $\mathbf{\Phi}$. In general, addressing this type of optimization problem can be quite challenging. To tackle this issue, we restructure the quartic functions into the SOS form. Subsequently, we employ the SDR algorithm to convert this high-order optimization problem into a quadratic optimization problem. Let  $\mathbf{v}=\left[v_{1}, \ldots, v_{N}\right]^{\rm{H}}$ denote the collection of the conjunction of the diagonal elements of $ \mathbf{\Phi} $. Upon introducing an auxiliary variable $t$ with $t^2 = 1$, we create a vector $\overline{\mathbf{v}} = [\mathbf{v}^{\rm{H}}, t^{*}]^{\rm{H}}$, and its corresponding covariance matrix is defined as $\overline{\mathbf{V}} = \overline{\mathbf{v}} \overline{\mathbf{v}}^{\rm{H}}$.
	
	In the following, we convert the objective function and constraints in (\ref{optimize}) into linear and quadratic forms of $\overline{\mathbf{V}}$, and then we solve the problem using the SDR algorithm.

	\subsubsection {Radar SINR Metric} \
	
	By using Lemma 1, we obtain the lower bound of the radar SINR at the $i$-th iteration as			
	\begin{equation}  \label{sbsn}
		\begin{aligned} 
			&\operatorname{Tr}\left(\mathbf{BRB}^{\rm{H}}\mathbf{J}^{-1}\right) 
			\geq 2\operatorname{Re}(\operatorname{Tr}\left(\mathbf{B}\mathbf{R}\mathbf{B}^{\rm{H}}_i\mathbf{J}^{-1}_i\right)) \\
			&-\operatorname{Tr}\left(\mathbf{J}^{-1}_i \mathbf{B}_i \mathbf{R}\mathbf{B}^{\rm{H}}_i\mathbf{J}^{-1}_i\mathbf{J}\right)  .
		\end{aligned}
	\end{equation} 
	Using property $	\operatorname{Tr}\left(\mathbf{A}^{\mathrm{H}} \mathbf{B}\right)=(\operatorname{vec}(\mathbf{A}))^{\mathrm{H}  }(\operatorname{vec}(\mathbf{B}))$, the first term on the right hand side of (\ref{sbsn}) is reformulated as
	\begin{equation} \label{bsn}
		\begin{aligned}
			&\operatorname{Tr}\left(\mathbf{B}\mathbf{R}\mathbf{B}^{\rm{H}}_i\mathbf{J}^{-1}_i\right) 
			=\operatorname{Tr}\left({\mathbf{\Phi }} ^{\mathrm{H}}{\mathbf{A\Phi G} \mathbf{R}\mathbf{B}^{\mathrm{H}}_i\mathbf{J}^{-1}_i{\mathbf{G}}^{\mathrm{H}} }\right) \\
			&=\operatorname{vec}({\mathbf{\Phi A\Phi}^{\rm{H}}})^{\mathrm{H}}\operatorname{vec}({\mathbf{ G} \mathbf{R}\mathbf{B}^{\mathrm{H}}_i\mathbf{J}^{-1}_i{\mathbf{G}}^{\mathrm{H}} }). 		
		\end{aligned}
	\end{equation}
	Since $\mathbf{\Phi}$ is a diagonal matrix,	we have
	\begin{equation}
		\begin{aligned}
			\mathbf{\Phi A \Phi}^{\rm{H}}=\mathbf{A} \odot \mathbf{vv}^{\rm{H}} 
			=\mathbf{A} \odot [\overline{\mathbf{V}}]_{1: N, 1: N}.
		\end{aligned}
	\end{equation}
	According to \cite[Eq. (1.11.15)]{zhang2017matrix}, one obtains
	\begin{equation}  \label{bgdrf}
		\begin{aligned}
			\operatorname{vec}(\mathbf{\Phi^{\rm{H}} A \Phi} )&=\operatorname{vec}(\mathbf{A}^\mathrm{H}  \odot [\overline{\mathbf{V}}]_{1: N, 1: N} ) \\
			&=\operatorname{vec}(\mathbf{A}^\mathrm{H}) \odot  \operatorname{vec}([\overline{\mathbf{V}}]_{1: N, 1: N}) \\
			&=\operatorname{Diag}(\operatorname{vec}(\mathbf{A}^\mathrm{H}))\operatorname{vec}([\overline{\mathbf{V}}]_{1: N, 1: N}).
		\end{aligned}
	\end{equation}
	By defining $\hat{\mathbf{v}}\triangleq\operatorname{vec}([\overline{\mathbf{V}}]_{1: N, 1: N})$ and $\mathbf{n}_1 \triangleq \operatorname{Diag}(\operatorname{vec}(\mathbf{A}^\mathrm{H}))^{\rm{H}}\operatorname{vec}({\mathbf{ G} \mathbf{R}\mathbf{B}^{\rm{H}}_i\mathbf{J}^{-1}_i{\mathbf{G}}^{\rm{H} }})$,  (\ref{bsn}) can be rewritten as 
	\begin{align}
		\operatorname{Tr}\left(\mathbf{B}\mathbf{R}\mathbf{B}^{\rm{H}}_i\mathbf{J}^{-1}_i\right)&=\operatorname{vec}([\overline{\mathbf{V}}]_{1: N, 1: N})^{\rm{H}}\operatorname{Diag}(\operatorname{vec}(\mathbf{A}^\mathrm{H}))^{\rm{H}} \notag\\
		&\operatorname{vec}({\mathbf{ G} \mathbf{R}\mathbf{B}_i^{\rm{H}}\mathbf{J}^{-1}_i{\mathbf{G}}^{\rm{H}} })  
		=\hat{\mathbf{v}}^{\rm{H}}\mathbf{n}_1.  \label{hsh}
	\end{align}
	Based on the above derivations, we transform the quadratic function of $\mathbf{\Phi}$ into a linear function of $\overline{\mathbf{V}}$.
	
	Defining $ \mathbf{T}_i\triangleq\mathbf{J}^{-1}_i  \mathbf{B}_i \mathbf{R}\mathbf{B}_i^{\rm{H}}\mathbf{J}^{-1}_i \succeq 0 $, we have
	{\color{black} 
		\begin{align} \label{what the fuck}
			&\operatorname{Tr}\left(\mathbf{J}^{-1}_i \mathbf{B}_i  \mathbf{R}\mathbf{B}_i^{\rm{H}}\mathbf{J}^{-1}_i \mathbf{J}\right)=\operatorname{Tr}(\mathbf{T}_i\mathbf{J})\nonumber \\
			&=\sigma^2\operatorname{Tr}(\mathbf{T}_i\mathbf{CC} ^{\rm{H}})+\eta \operatorname{Tr}(\mathbf{T}_i\mathbf{G} ^{\rm{H}} \mathbf{\Phi G R G}^{\rm{H}} \mathbf{\Phi}^{\rm{H}}\mathbf{G})\nonumber \\			
			&\quad +2\sigma^2\operatorname{Tr}(\mathbf{T}_i\mathbf{G} ^{\rm{H}}\mathbf{\Phi}
			^{\rm{H}}\mathbf{\Phi G} )+2\sigma^2\operatorname{Re}(\operatorname{Tr}({\mathbf{T}_i{\mathbf{G}}^{\rm{H}}{\mathbf{\Phi }}{\mathbf{\Phi }} ^{\rm{H}}{\mathbf{A}}{\mathbf{\Phi G}}}))\nonumber \\
			&\quad +\underbrace{\sigma_{\rm{r}}^2\operatorname{Tr}(\mathbf{T}_i)}_\textrm{const}.
		\end{align}
	} 
	By using the equality $\operatorname{Tr}(\mathbf{Q} \mathbf{P} \mathbf{S} \mathbf{K})=\left(\operatorname{vec}\left(\mathbf{K}^{\mathrm{T}}\right)\right)^{\mathrm{T}}\left(\mathbf{S}^{\mathrm{T}} \otimes \mathbf{Q}\right) \operatorname{vec}(\mathbf{P})$, 
	the first term in (\ref{what the fuck}) can be transformed as
	\begin{align} 
		&\sigma^2{\operatorname{Tr}}\left(\mathbf{T}_i{\mathbf{G}}^{\rm{H}}{\mathbf{\Phi }}^{\rm{H}}{\mathbf{A\Phi}}\mathbf{\Phi}^{\rm{H}} \mathbf{A}^{\rm{H}}\mathbf{\Phi G}   \right) \notag \\
		&=\sigma^2{\operatorname{Tr}}\left({\mathbf{G}\mathbf{T} _i\mathbf{G}}^{\rm{H}}{\mathbf{\Phi }}^{\rm{H}}{\mathbf{A\Phi}}{\mathbf{I}}_N\mathbf{\Phi}^{\rm{H}} \mathbf{A}^{\rm{H}}\mathbf{\Phi }   \right)  \notag \\
		&=\sigma^2\operatorname{vec}\left(\mathbf{\Phi}^{\rm{H}} \mathbf{{A \Phi}} \right)^{\rm{H}}\left(\mathbf{I}_N\otimes\left(\mathbf{G}\mathbf{T}_i\mathbf{G}^{\rm{H}}\right)\right) \operatorname{vec}\left(\mathbf{\Phi}^{\rm{H}} \mathbf{{A \Phi}} \right) \notag \\
		&=\sigma^2\operatorname{vec}([\overline{\mathbf{V}}]_{1: N, 1: N})^{\rm{H}}\operatorname{Diag}(\operatorname{vec}(\mathbf{A}^\mathrm{H}))^{\rm{H}}\left(\mathbf{I}_{N} \otimes \left(\mathbf{G}\mathbf{T}_i\mathbf{G}^{\rm{H}}\right)\right) \notag \\
		&\operatorname{Diag}(\operatorname{vec}(\mathbf{A}^\mathrm{H}))\operatorname{vec}([\overline{\mathbf{V}}]_{1: N, 1: N}).  \label{etbs}
	\end{align}		
	It can be verified that the matrix $\mathbf{I}_{N} \otimes \left(\mathbf{G}\mathbf{T}_i\mathbf{G}^{\rm{H}}\right)$ is a positive semidefinite matrix. For a positive semidefinite matrix, we apply the Cholesky decomposition as $\mathbf{I}_{N} \otimes \left(\mathbf{G}\mathbf{T}_i\mathbf{G}^{\rm{H}}\right)=\mathbf{L}_1\mathbf{L}_1^{\rm{H}}$. 
	Thus, we define $\mathbf{M}_1\triangleq\sigma\mathbf{L}_1\operatorname{Diag}(\operatorname{vec}(\mathbf{A}^\mathrm{H}))$, and (\ref{etbs}) is now transformed into a quadratic function of $\overline{\mathbf{V}}$:
	\begin{equation}  \label{gf}
		\sigma^2{\operatorname{Tr}}\left(\mathbf{T}_i{\mathbf{G}}^{\rm{H}}{\mathbf{\Phi }}^{\rm{H}}{\mathbf{A\Phi}}\mathbf{\Phi}^{\rm{H}} \mathbf{A}^{\rm{H}}\mathbf{\Phi G}   \right)=\|\mathbf{M}_1\hat{\mathbf{v}}\|_2^2.
	\end{equation}	
	The second term of (\ref{what the fuck}) can be reformulated as:
	\begin{align} \label{bsg}
		&\eta {\operatorname{Tr}}\left(\mathbf{T}_i\mathbf{\mathbf{G} ^{\rm{H}} \mathbf{\Phi G R G}^{\rm{H}} \mathbf{\Phi}^{\rm{H}}\mathbf{G}}\right)\notag \\
		&=\eta (\operatorname{vec}(\mathbf{GRG}^{\rm{H}} ))^{\rm{H}} \operatorname{vec}(\mathbf{\Phi}^{\rm{H}}\mathbf{G}\mathbf{T}_i\mathbf{G}^{\rm{H}} \mathbf{\Phi}  ) \notag \\
		&=\eta (\operatorname{vec}(\mathbf{GRG}^{\rm{H}} ))^{\rm{H}} \operatorname{Diag}(\operatorname{vec}(\mathbf{G}\mathbf{T}_i\mathbf{G}^{\rm{H}}))\operatorname{vec}([\overline{\mathbf{V}}]_{1: N, 1: N}) \notag \\
		&\triangleq\mathbf{n}_2^{\rm{H}}\hat{\mathbf{v}},
	\end{align}
	where $\mathbf{n}_2^{\rm{H}}=\eta(\operatorname{vec}(\mathbf{GRG}^{\rm{H}} ))^{\rm{H}} \operatorname{Diag}(\operatorname{vec}(\mathbf{G}\mathbf{T}_i\mathbf{G}^{\rm{H}}))$.
	Note that for the diagonal matrix $\mathbf{\Phi}\mathbf{\Phi}^{\rm{H}}$, we have
	\begin{equation}
		\begin{aligned}
			\mathbf{\Phi \Phi}^{\rm{H}}
			&=\operatorname{Diag}\left(\operatorname{diag}\left([\overline{\mathbf{V}}]_{1: N, 1: N}\right)\right) \triangleq \hat{\mathbf{V}},		
		\end{aligned}
	\end{equation}
	hence, the third term of (\ref{what the fuck}) can be rewritten as
	\begin{equation}
		\begin{aligned} \label{dzf}
			2\sigma^2{\operatorname{Tr}}\left(\mathbf{T}_i\mathbf{G} ^{\rm{H}}\mathbf{\Phi} ^{\rm{H}}\mathbf{\Phi G}\right)&=2\sigma^2{\operatorname{Tr}}\left(\mathbf{T}_i\mathbf{G} ^{\rm{H}}\hat{\mathbf{V}}\mathbf{G}\right). \\
		\end{aligned}
	\end{equation}

	Note that the fourth term of (\ref{what the fuck}) is a cubic function of $\mathbf{v}$, which can not directly transformed into a function of $\overline{\mathbf{V}}$. We approximate it according to the following Lemma:
	
	{\color{black}	\itshape \textbf{Lemma 3:} \upshape Define $\mathbf{K}_i$ and $\mathbf{L}_i$ as the matrix $\mathbf{K}$ and $\mathbf{L}$ at the  $i$-th iteration, we have
		\begin{equation}
			2\operatorname{Re}(\operatorname{Tr}(\mathbf{KL}^{\rm{H}}))\le {\frac{\|\mathbf{L}_i\|_F}{\|\mathbf{K}_i\|_F}}\operatorname{Tr}(\mathbf{KK}^{\rm{H}})+{\frac{\|\mathbf{K}_i\|_F}{\|\mathbf{L}_i\|_F}}\operatorname{Tr}(\mathbf{LL}^{\rm{H}}).
	\end{equation} }
	{\color{black}
		\itshape \textbf{Proof:}  \upshape This inequality can be obtained by expanding the expression  $\left\|\sqrt{\frac{\|\mathbf{L}_i\|_F}{\|\mathbf{K}_i\|_F}}\mathbf{K}-\sqrt{\frac{\|\mathbf{K}_i\|_F}{\|\mathbf{L}_i\|_F}}\mathbf{L}\right\|^2_F \geq 0$. $\hfill\blacksquare$}
	
	{\color{black}
		According to Lemma 3 and the MM framework, the fourth term of (\ref{what the fuck}) can be upper bounded by
		\begin{equation} \label{bfd}
			\begin{aligned}
				&2\sigma^2\operatorname{Re}(\operatorname{Tr}({\mathbf{T}_i{\mathbf{G}}^{\rm{H}}{\mathbf{\Phi }}{\mathbf{\Phi }} ^{\rm{H}}{\mathbf{A}}{\mathbf{\Phi G}}}))  \\
				& \leq {\sigma^2}\left(\alpha^2\|\mathbf{GT}_i\mathbf{G}^{\rm{H}}\mathbf{ \Phi  }\|^2_F +\frac{1}{\alpha^2 }\|\mathbf{ \Phi}^{\rm{H}}\mathbf{A \Phi }\|^2_F\right),
			\end{aligned}
		\end{equation} 
		where $\alpha^2=\frac{\|\mathbf{ \Phi}_i^{\rm{H}}\mathbf{A \Phi }_i\|_F}{\|\mathbf{GT}_i\mathbf{G}^{\rm{H}}\mathbf{ \Phi  }_i\|_F}$.}
	
	{\color{black}
		By adopting the same method in (\ref{etbs}) and (\ref{bsg}), the right hand side of (\ref{bfd}) can be written as 
		\begin{equation}
			\begin{aligned} \label{fbbg}
				&{\sigma^2}\left(\alpha^2\|\mathbf{GT}_i\mathbf{G}^{\rm{H}}\mathbf{ \Phi  }\|^2_F +\frac{1}{\alpha^2 }\|\mathbf{ \Phi}^{\rm{H}}\mathbf{A \Phi }\|^2_F\right)\\
				&=\operatorname{Tr}(\mathbf{K}_i\hat{\mathbf{V}})+\|\mathbf{M}_2\hat{\mathbf{v}}\|_2^2,
			\end{aligned}
		\end{equation}
		where $\mathbf{K}_i\triangleq\sigma^2\alpha^2\mathbf{G}\mathbf{T}_i\mathbf{G}^{\rm{H}}\mathbf{GT}_i\mathbf{G}^{\rm{H}}$, and $\mathbf{M}_2\triangleq\frac{\sigma}{\alpha} \operatorname{Diag}(\operatorname{vec}(\mathbf{A}^\mathrm{H}))$.}
	
	Combining (\ref{bgdrf}), (\ref{gf}), (\ref{bsg}), (\ref{dzf}) and (\ref{fbbg}) together, the overall radar SINR expression can then be transformed into
	\begin{equation}
		\begin{aligned}
			\rm{SINR}&=2\operatorname{Re}(\hat{\mathbf{v}}^{\rm{H}}\mathbf{n}_1)-\|\mathbf{M}_1\hat{\mathbf{v}}\|_2^2-\mathbf{n}_2^{\rm{H}}\hat{\mathbf{v}} \label{zbfd}\\
			&-2\sigma^2{\operatorname{Tr}}\left(\mathbf{T}_i\mathbf{G} ^{\rm{H}}\hat{\mathbf{V}}\mathbf{G}\right)-\|\mathbf{M}_2\hat{\mathbf{v}}\|_2^2-\operatorname{Tr}(\mathbf{K}_i\hat{\mathbf{V}}).
		\end{aligned}
	\end{equation}

	\subsubsection {Active RIS Transmit Power Constraint} \
	
	In the following, we transform the transmit power of the active RIS in (\ref{RIS power}) into a quadratic function of $\overline{\mathbf{V}}$.
	The first term of (\ref{RIS power}) can be expressed as
	\begin{equation}
		{\operatorname{Tr}}(\mathbf{\Phi^{\rm{H}} A \Phi G R G^{\rm{H}}  \Phi^{\rm{H}} A^{\rm{H}} \Phi})
		\triangleq \|\mathbf{M}_3\hat{\mathbf{v}}\|_2^2,		
	\end{equation} 
	where $\mathbf{M}_3=\mathbf{L}_2\operatorname{Diag}(\operatorname{vec}(\mathbf{A}^\mathrm{H}))$ and the positive semidefinite matrix $\mathbf{I}_{N} \otimes \left(\mathbf{G}\mathbf{R}\mathbf{G}^{\rm{H}}\right) =\mathbf{L}_2\mathbf{L}_2^{\rm{H}}$. 
	The second term of (\ref{RIS constraint}) can be expressed as
	\begin{equation}
		\begin{aligned}
			\sigma^2 {\operatorname{Tr}}(\mathbf{\Phi^{\rm{H}} A \Phi \Phi^{\rm{H}} A^{\rm{H}} \Phi})
			&\triangleq \|\mathbf{M}_4\hat{\mathbf{v}}\|_2^2,
		\end{aligned}
	\end{equation}
	where $\mathbf{M}_4=\sigma \operatorname{Diag}(\operatorname{vec}(\mathbf{A}^\mathrm{H}))$.
	The rest terms of (\ref{RIS constraint}) can be expressed as		
	\begin{equation}
		\begin{aligned}
			{\operatorname{Tr}}(\mathbf{\Phi G R G^{\rm{H}} \Phi^{\rm{H}}} )={\operatorname{Tr}}(\mathbf{ G R G}^{\rm{H}} \hat{\mathbf{V}}),
		\end{aligned}
	\end{equation}
	and
	\begin{equation}
		\begin{aligned}
			{\operatorname{Tr}}(\mathbf{\Phi ^{\rm{H}}\Phi })=\operatorname{Tr}(\hat{\mathbf{V}}).
		\end{aligned}
	\end{equation}
	
	Therefore, the active RIS transmit power constraint can be expressed as
	\begin{equation} \label{zv}
		\|\mathbf{M}_3\hat{\mathbf{v}}\|_2^2+\|\mathbf{M}_4\hat{\mathbf{v}}\|_2^2+{\operatorname{Tr}}(\mathbf{ G R G}^{\rm{H}} \hat{\mathbf{V}})+2\sigma^2\operatorname{Tr}(\hat{\mathbf{V}})-P_{\rm{RIS}} \le 0.
	\end{equation}
	
	\subsubsection {Communication Metric and Power Amplification Gain Constraint} \
	
	Defining $\mathbf{H}_k=[\mathbf{G}^{\rm{H}}\operatorname{Diag}\left(\mathbf{h}_{1, k}\right), \, \mathbf{h}_{2,k}]^{\rm{H}}$, the equivalent channel between the BS and UE $k$ can be written as 
	${\mathbf{h}}_k^{\rm{H}}=\overline{\mathbf{v}}^{\rm{H}}\mathbf{H}_k$. 
	Besides, the remaining term $ {d}_k $ can be formulated as $	d_k={\mathbf{h}}_{1,k}^{\rm{H}}\operatorname{Tr}(\hat{\mathbf{V}}){\mathbf{h}}_{1,k}\sigma^{2}+\sigma _{\rm{z}}^{2}$.
	Thus, the communication QoS constraints can be formulated as
	\begin{equation} \label{bdfb}
		\begin{aligned}
			&(1+{\mathbf{\xi}}^{-1})\operatorname{Tr}(\overline{\mathbf{R}}_{2,k}\overline{\mathbf{V}}) -\operatorname{Tr}(\overline{\mathbf{R}}_{1,k}\overline{\mathbf{V}}) \ge \\ &\sigma^{2}\operatorname{Tr}(\hat{\mathbf{V}}{\mathbf{h}}_{1,k}{\mathbf{h}}_{1,k}^{\rm{H}})  )+\sigma _{\rm{z}}^{2},\quad k=1,\cdots,K,
		\end{aligned}
	\end{equation}
	where $\overline{\mathbf{R}}_{1,k}$ and $\overline{\mathbf{R}}_{2,k}$ are defined as $\overline{\mathbf{R}}_{1,k}\triangleq\mathbf{H}_k\mathbf{R}\mathbf{H}_k^{\rm{H}}$ and $\overline{\mathbf{R}}_{2,k}\triangleq\mathbf{H}_k\mathbf{R}_k\mathbf{H}_k^{\rm{H}}$, respectively.
	
	Finally, the power amplification gain constraint of the active RIS $|v_n|^2\le a_{\rm{RIS}} , n = 1, \cdots ,N,$ can be formulated as  
	\begin{equation} \label{bdf}
		[\operatorname{diag}(\overline{\mathbf{V}})]_{n} \leq a_{\rm{RIS}},\quad n=1,2,\cdots,N.
	\end{equation}

	\subsubsection {Problem Reformulation and the Proposed Algorithm}  \

	Following the aforementioned transformation, the active RIS reflecting coefficients optimization problem can be reformulated as:
	\begin{subequations}\label{optimize for precoding4}
		\begin{align}
			\mathop {\max }\limits_{{\overline{\mathbf{V}} } }  \quad & (\ref{zbfd}) \label{whattttt}\\
			\qquad\ \textrm{s.t.} \quad
			&(\ref{zv}), (\ref{bdfb}), (\ref{bdf}),  \label{jiuzhezhe} \\	
			&[\operatorname{diag}(\overline{\mathbf{V}})]_{N+1} = 1, \label{rbgb}\\
			&\rm{Rank}(\overline{\mathbf{V}})=1.\label{befab} 					
		\end{align}
	\end{subequations}
	Recall that $\hat{\mathbf{V}}=\operatorname{Diag}\left(\operatorname{diag}\left([\overline{\mathbf{V}}]_{1: N, 1: N}\right)\right) $ and $\hat{\mathbf{v}}=\operatorname{vec}([\overline{\mathbf{V}}]_{1: N, 1: N})$ are affine functions of the optimization variable $\overline{\mathbf{V}}$, which will not affect the curvature. The overall problem can be regarded as an SDP problem when constraint (\ref{befab}) is relaxed.
	After the solution to the relaxed problem $\overline{\mathbf{V}}^{\star}$ is obtained using CVX tools, we apply the Gaussian randomization method or eigenvalue decomposition method to construct the rank-one solution $\mathbf{v}^{\star}$. 
	
	The details of the proposed algorithm for solving Problem (\ref{optimize}) is summarized in Algorithm 1.
	
	\begin{algorithm}
		\caption{Joint beamforming for the active RIS-aided ISAC system} 
		\begin{algorithmic}
			\STATE \textbf{Input}: The maximum iteration time for the beamforming matrix optimization $t^{\rm{max}}_1$, the maximum iteration time for the reflecting coefficient matrix optimization $t_2^{\rm{max}}$, the maximum iteration time for the AO $t^{\rm{max}}$.
		\end{algorithmic}	
		\begin{algorithmic}[1]
			\STATE Set $t=0$, generate a random reflecting coefficient $\mathbf{v}^0$.
			\STATE Drop the rank-one constraint and obtain a feasible solution of the relaxed Problem (\ref{optimize for precoding7}) $\tilde{\mathbf{R}}_{k}$ and $\tilde{\mathbf{R}}$ via CVX.
			\STATE Calculate the beamforming vector $\hat{\mathbf{w}}_{{\rm{c}},k} $ and matrix $\hat{\mathbf{W}}_{\rm{r}} $ according to (\ref{wck}) and  (\ref{wr}), and set $\mathbf{W}^{0}=[\hat{\mathbf{W}_{\mathrm{r}}}, \hat{\mathbf{w}}_{{\rm{c}},1},\hat{\mathbf{w}}_{{\rm{c}},2},......,\hat{\mathbf{w}}_{{\rm{c}},K} ]$.
			\STATE  \textbf{Repeat} 
			\STATE \quad Set $t_1=0$, and $\mathbf{W}_{0}=\mathbf{W}^t$.
			\STATE \quad \textbf{Repeat} 
			\STATE \quad \quad  Obtain ${\mathbf{W}}_{t_1+1}$ by solving (\ref{optimize for precoding2}) with given ${\mathbf{W}}_{t_1}$.
			\STATE \quad \quad $ t_1=t_1+1 $.
			\STATE \quad \textbf{Until $t_1=t_{1}^{\rm{max}}$}
			\STATE \quad Set $t_2=0$,  and $\mathbf{v}_0=\mathbf{v}^{t}$.
			\STATE \quad \textbf{Repeat} 
			\STATE \quad \quad   Obtain ${\mathbf{v}}_{t_2+1}$ by solving Problem (\ref{optimize for precoding4}) with fixed ${\mathbf{v}}_{t_2}$.
			\STATE \quad \quad $ t_2=t_2+1 $.
			\STATE \quad \textbf{Until $t_2=t_{2}^{\rm{max}}$}
			\STATE \quad $\mathbf{W}^t=\mathbf{W}_{t_1}$, $\mathbf{v}^t=\mathbf{v}_{t_2}$,  and $t=t+1$.
			\STATE  \textbf{Until $t=t^{\rm{max}}$}	
		\end{algorithmic}
	\end{algorithm}

	\vspace{-0.5cm}
	\subsection{{Computational Complexity of the Proposed Algorithm }}
	The computational complexity of solving Problem (\ref{optimize for precoding2}) and Problem (\ref{optimize for precoding4}) mainly lies in the interior point method and is given by\cite{53}
	\begin{equation}
		{\footnotesize 
			\mathcal{O}\left( \underbrace{\left(\sum_{j=1}^{J} k_{j}+2 m\right)^{1 / 2}}_{\text {Iteration Complexity }}n(n^{2}+\underbrace{n \sum_{j=1}^{J} k_{j}^{2}+\sum_{j=1}^{J} k_{j}^{3}}_{\text {due to LMI }}+\underbrace{ \sum_{i=1}^{m} a_{i}^{2}}_{\text {due to SOC }})\right), }
	\end{equation}
	where $n$ denotes the number of variables, $J$ denotes the number of the linear matrix inequality (LMI) constraints, $k_{j}$ is the dimension of the $j$-th LMI constraint, $m$ denotes the number of the second-order cone (SOC) constraints, and $a_i$ is the dimension of the $i$-th SOC constraint.
	
	For Problem (\ref{optimize for precoding2}), the number of the variables  is $n_1=M(M+K)$, and it contains $3$ SOC constraints of size $M(M+K)$ and $K$ SOC constraints with the size $M$. {\color{black}We ignore the constant value, and the approximate computation complexity is given by $o_{{f}}=\mathcal{O}\left( {\left(2K\right)^{1 / 2}}n_1(n_1^{2}+{ (M(M+K))^{2}})+Km^2\right)=\mathcal{O}\left(K ^{3.5}  {M}^{6}  \right)$. Similarly, the approximate computational complexity of solving Problem (\ref{optimize for precoding4}) is given by $o_{{e}}=\mathcal{O}\left( {K^{1.5} N^6}\right)$. Given that the number of reflecting elements $N$ significantly exceeds the number of antennas $M$, the main computational complexity lies in optimizing the reflecting coefficients of the RIS.} Defining $t^{\rm{max}}$ as the number of iterations of the AO algorithm, $t_1^{\rm{max}}$ as the number of iterations of beamforming matrix optimization, and $t_2^{\rm{max}}$ as the number of iterations of reflecting coefficient matrix optimization, the proposed algorithm's computational complexity, is $t^{\rm{max}}(t_1^{\rm{max}}o_f+t_2^{\rm{max}}o_e)$. 
	
	{\color{black} \textit{Remarks:} It is worth noting that the existing studies on active RIS also exhibit a similar level of complexity \cite{47}. Besides, the algorithm in \cite{46} may not guarantee a consistently increasing and convergence performance when optimizing the sensing capability w.r.t. the number of iterations.}
	

	{\color{black}
		\section{Performance analysis of the radar SINR in active RIS-aided sensing systems}
		In the preceding sections, we focused on configuring $\mathbf{W}$ and $\mathbf{\Phi}$ through the development of the algorithm. Now, we shift our attention to examining the impact of several key factors on the radar SINR. Specifically, we investigate how the number of reflecting elements in the active RIS, the transmit  power allocation between the DFRC BS and the active RIS, and the deployment of the active RIS influence the radar SINR.
		
		\subsection{Simplified System Setting}
		We consider a single antenna radar with the aid of an active RIS equipped with $N$ elements, which only performs sensing function. The transmit power of the radar is given by $P_t$. The LoS channel is considered to characterize the channel between the single antenna radar and the active RIS, i.e., $\mathbf{g}=g {\mathbf{a}}_2(\theta_2)$, where $g$ is the distance-dependent path loss factor of $\mathbf{g}$. Unlike the previously mentioned models, where each reflecting element has an individual amplitude, we assume that the active RIS is equipped with a shared amplifier.  Consequently, the reflecting coefficients matrix $\mathbf{\Phi}$ is represented as $\mathbf{\Phi}=\varrho\mathbf{\Psi}$, with $\mathbf{\Psi}=\operatorname{Diag}{(\left[ \psi_1,\cdots,\psi_n,\cdots,\psi_N \right])}$ satisfying the equality $|\psi_i|=1$ for $i=1,\cdots, N$, and the power amplification gain $\varrho^2$ should no greater than $a_{\rm{RIS}}=40$ dB. 
		
		We assume that the radar is located at $(0, 0)$, the active RIS is located at $(x, d)$, and the sensing target is located at $(D, 0)$.	As for a realistic scenario, we assume the distance between the DFRC BS and the active RIS is not very small, i.e., $g^2 \le -70$ dB, and the round-trip path loss of the RIS-Target-RIS link $\beta_{\mathrm{r}}^2$ is no less than $-120$ dB. We also assume that the noise of the active RIS and the DFRC BS has the same noise power $\sigma_{\rm{r}}^2=\sigma^2=-100 $ dBm.
		
		Under the simplified system setting, the radar SINR in (\ref{SINR}) reduces to
		\begin{align} \label{bdgf}
			{\rm{ SINR}}&=\frac{ {P}_t \varrho^4 |{\mathbf{g}}^{\rm{H}}{\mathbf{\Psi }} ^{\rm{H}}{\mathbf{A\Psi g }}|^2}{\sigma^2_{\rm{r}}+\varrho ^4\sigma^2\|{\mathbf{g}}^{\rm{H}}{\mathbf{\Psi }} ^{\rm{H}}{\mathbf{A\Psi}}\|^2_2+2\varrho ^2\sigma^2\|{\mathbf{g}}^{\rm{H}}{\mathbf{\Psi }}\|^2_2+g_1} \notag\\
			&=\frac{ {P}_t \varrho^4 f^4}{\sigma^2_{\rm{r}}+\varrho ^4\sigma^2\|{f\sqrt{\beta_{\mathrm{r}}}\mathbf{a}_{3}^{\rm{H}}(\theta_3)\mathbf{\Psi}}\|^2_2+2\varrho ^2\sigma^2\|{\mathbf{g}}^{\rm{H}}{\mathbf{\Psi }}\|^2_2+g_1} \notag \\
			&=\frac{ {P}_t \varrho^4 f^4}{\sigma^2_{\rm{r}}+\varrho ^4\sigma^2 f^2\|\sqrt{\beta_{\mathrm{r}}}{\mathbf{a}_{3}^{\rm{H}}(\theta_3)\|^2_2+2\varrho ^2\sigma^2\|{\mathbf{g}}^{\rm{H}}\|^2_2+g_1}} \notag \\
			&=\frac{ {P}_t \varrho^4 f^4}{\sigma^2_{\rm{r}}+\varrho ^4\sigma^2 f^2 N \beta_{\mathrm{r}}+2\varrho ^2\sigma^2 N g^2+g_1},
		\end{align}
		where $g_1\triangleq2\varrho^3\operatorname{Re}(\sigma^2{\mathbf{g}}^{\rm{H}}{\mathbf{\Psi \Psi}}^{\rm{H}}{\mathbf{A }}\mathbf{\Psi g})+\eta^2 \varrho^2 {P}_t |{\mathbf{g}}^{\rm{H}}{\mathbf{\Psi }}{\mathbf{g}}|^2$, and $f\triangleq|\sqrt{\beta_{\mathrm{r}}} {\mathbf{g}}^{\rm{H}}{\mathbf{\Psi }} ^{\rm{H}}{\mathbf{a}_{3}(\theta_3)}|$.
		
		The transmit power constraint in (\ref{RIS constraint}) can be written as:		
		\begin{align} 
			&{P}_t\varrho^4{\operatorname{Tr}}(\mathbf{ A \Psi g  g^{\rm{H}} \Psi^{\rm{H}} A^{\rm{H}} })+\sigma^2\varrho^4 \|\mathbf{ A  }\|_F^2 
			+{P}_t\varrho^2 \|\mathbf{ g  }\|_2^2 +2\sigma^2\varrho^2 \notag \\
			&={P}_t\varrho^4f^2\|\sqrt{\beta_{\mathrm{r}}}\mathbf{a}_{3}(\theta_3)\|^2_2+\sigma^2\varrho^4 \|\mathbf{ A  }\|_F^2
			+{P}_t\varrho^2 \|\mathbf{ g  }\|_2^2 +2\sigma^2\varrho^2 \notag \\	
			&={P}_t\varrho^4f^2 N \beta_{\mathrm{r}}+\sigma^2\varrho^4 N^2 \beta_{\mathrm{r}}^2 
			+{P}_t\varrho^2 N g^2 +2\sigma^2\varrho^2  \notag \\			
			&\le P_{\rm{RIS}}. \label{RIS constraintsvd}
		\end{align}
		
		In most cases, the power of the reflected signal is much higher than the power of the introduced thermal noise at the active RIS. The transmit power constraint can be simplified as  ${P}_t\varrho^4 f^2 N \beta_{\mathrm{r}}
		+{P}_t\varrho^2 N g^2 			
		\le P_{\rm{RIS}}$.
		
		Thus, the optimization problem in the simplified system can be formulated as
			%
		\begin{subequations}\label{optimizve}
			\begin{align}
				\mathop {\max }\limits_{{{\varrho,f^2 } } } \quad  &\frac{ {P}_t \varrho^4 f^4}{\sigma^2_{\rm{r}}+\varrho ^4\sigma^2 f^2 N \beta_{\mathrm{r}}+2\varrho ^2\sigma^2 N g^2+g_1} \label{bfgvb} 
				\\
				\textrm{s.t.}\quad &\varrho^2\le a_{\rm{RIS}},  \label{dbs}\\
				&{P}_t\varrho^4 f^2 N \beta_{\mathrm{r}}+{P}_t\varrho^2 N g^2 \le P_{\rm{RIS}}. \label{vbxz}
			\end{align}
		\end{subequations}
		The optimization variables $f^2$ and $\varrho$ respectively represent the beamforming gain and the power amplification gain induced by the active RIS. It is imperative that at least one of the inequality constraints (\ref{dbs}) or (\ref{vbxz}) is transformed into an equality constraint to maximize the SINR. This observation leads us to consider two cases: a) the transmit power budget of the active RIS is insufficient, i.e., ${P}_t\varrho^4 f^2 N \beta_{\mathrm{r}}
		+{P}_t\varrho^2 N g^2 			
		= P_{\rm{RIS}}$, and b) the power amplification gain is limited, i.e., $\varrho^2=a_{\rm{RIS}}$.

		\subsection{Power Scaling Law Analysis}	
		Since Problem (\ref{optimizve}) is still challenging to solve, we cannot directly obtain the asymptotic radar SINR. An alternative solution is to consider some approximation of the radar SINR. Specifically, we first derive the scaling order of both the lower bound and the upper bound of the radar SINR, then the scaling order of the radar SINR can be obtained according to the Sandwich Theorem.   
		
		\itshape \textbf{Lemma 4:} \upshape By only considering the thermal noise introduced by the first reflected signal, the upper bound of the radar SINR is given by
		\begin{equation}
			\begin{aligned} \label{fdbzdf}
				{\rm{ SINR}}\le {\rm{ SINR}}_{\rm{up}}\triangleq\frac{ {P}_t \varrho^4 f^4}{\varrho ^4\sigma^2 f^2 N \beta_{\mathrm{r}}} 
				=\frac{ {P}_t f^2}{\sigma^2  N \beta_{\mathrm{r}}},
			\end{aligned}
		\end{equation} 	
		which has a scaling order of $\mathcal{O}\left(N\right)$. 
		
		\itshape \textbf{Proof:}  \upshape ${\rm{ SINR}}_{\rm{up}}$ is a monotonically increasing function w.r.t. the beamforming gain $f^2$. Hence, we can always obtain the maximum beamforming gain while decreasing the power amplification gain to make sure constraint (\ref{dbs}) and (\ref{vbxz}) are satisfied.  The maximum $f^2=N^2 g^2 \beta_{\mathrm{r}}$ can be achieved by
		\begin{equation} \label{fvd}
			\psi_{n}=   \arg \left\{\left[\mathbf{a}_{3}(\theta_3)\right]_{n}\right\}-\arg \left\{\left[\mathbf{g}\right]_{n}\right\}, \forall n. 
		\end{equation}
		Thus, by letting $N \rightarrow \infty$, we can conclude that the upper bound of the radar SINR has a scaling order of $\mathcal{O}\left(N\right)$. $\hfill\blacksquare$

		When the number of the reflecting elements is large, a narrow beam is generated, the interference term $|{\mathbf{g}}^{\rm{H}}{\mathbf{\Psi }}{\mathbf{g}}|^2$ becomes negligible. Thus, we have  
	\begin{equation}
	\begin{aligned}
		g_1&=2\varrho^3\operatorname{Re}(\sigma^2{\mathbf{g}}^{\rm{H}}{\mathbf{\Psi \Psi}}^{\rm{H}}{\mathbf{A }}\mathbf{\Psi g})+\eta^2 \varrho^2 {P}_t |{\mathbf{g}}^{\rm{H}}{\mathbf{\Psi }}{\mathbf{g}}|^2 \\
		&\approx 2\varrho^3\operatorname{Re}(\sigma^2{\mathbf{g}}^{\rm{H}}{\mathbf{\Psi \Psi}}^{\rm{H}}{\mathbf{A }}\mathbf{\Psi g}) \\
		&<  2\varrho^3 \sigma^2 |\mathbf{g }^{\rm{H}}{\mathbf{a }_3(\theta_3)}|N {\beta_{\mathrm{r}}} g 
		<  2\varrho^3 \sigma^2 N^2 \beta_{\mathrm{r}} g^2\triangleq g_2.
	\end{aligned}
\end{equation}
		
		\itshape \textbf{Lemma 5:} \upshape The radar SINR can be lower bounded by 
		\begin{align} \label{gbd}
			{\rm{ SINR}} &\ge \frac{ {P}_t \varrho^4 f^4}{\sigma^2_{\rm{r}}+\varrho ^4\sigma^2 f^2 N \beta_{\mathrm{r}}+2\varrho ^2\sigma^2 N g^2+g_2} \notag \\
			&\ge \frac{{P}_t \varrho^4 N^4 g^4 \beta^2_{\mathrm{r}}}{\sigma^2_{\rm{r}}+\varrho ^4\sigma^2 N^2 g^2 \beta_{\mathrm{r}}^2 N+2\varrho ^2\sigma^2 N g^2+g_2} \notag \\
			&\triangleq{\rm{ SINR}}_{\rm{low}},
		\end{align}
		and the scaling order of ${\rm{ SINR}}_{\rm{low}}$ is $\mathcal{O}\left(N\right)$.
		
		\itshape \textbf{Proof:} \upshape If the power amplification gain is a constant, i.e., $\varrho^2=a_{\rm{RIS}}$, by letting $N \rightarrow \infty$, the lower bound of the radar SINR has a scaling order of $\mathcal{O}\left(N\right)$.
		
		When the transmit power of the active RIS is insufficient, the power amplification gain can be obtained by using the binary search algorithm according to (\ref{vbxz}). With a large number of reflecting elements, it becomes evident that the transmit power constraint can be further approximated as
		\begin{equation} \label{vzdf}
			\begin{aligned}
				{P}_t\varrho^4 N^3 \beta_{\mathrm{r}}^2 g^2 		
				\le P_{\rm{RIS}}.
			\end{aligned}
		\end{equation}
		This assumption holds true when the condition $N^2 \varrho^2\beta_{\mathrm{r}}^2 > 1$ is satisfied, which implies that the number of reflecting elements $N$ should be greater than $10^4$.
		
		By substituting (\ref{vzdf}) into (\ref{gbd}), it can be demonstrated that the lower bound also exhibits a scaling order of $\mathcal{O}\left(N\right)$. $\hfill\blacksquare$
		
		By integrating Lemma 4 and Lemma 5, and subsequently applying the Sandwich Theorem, we can determine the scaling order of the radar SINR, as detailed in the ensuing Theorem:
		
		\itshape \textbf{Theorem 1:} \upshape For the active RIS-aided sensing systems, the radar SINR increases with $N$ as $N \rightarrow \infty$ according to 
		\begin{equation}
			\lim _{N \rightarrow \infty} \frac{\log _{2}\rm{SINR}}{\log _{2} N}=1.
		\end{equation}
		
		\itshape \textbf{Remarks:} \upshape Theorem 1 shows that although the signal is reflected by the active RIS twice, the radar SINR only increases with a scaling order of $N$, as in the active RIS-aided communication systems\cite{21}. The scaling order can be intuitively explained as follows. The transmit sensing signal has experienced four-hop path loss, which in return, results in a scaling order of $\mathcal{O}\left(N^{4}\right)$. However, as indicated in the denominator of (\ref{fdbzdf}), the noise induced by the first reflected signal also experiences three-hop path loss, leading to a scaling order of $\mathcal{O}\left(N^{3}\right)$. 
		
		
		\subsection{Transmit Power Allocation for a Moderate Number of Reflecting Elements}
		In the realistic scenario, the number of the reflecting elements for the active RIS may not be infinitely large. It is hard for the amplification gain as well as the beamforming gain to fully compensate for the round-trip path loss of the RIS-Target-RIS link. When the  number of the reflecting elements is $16$ \cite{Active_RIS_model}, the power of the second reflected signal is much smaller than the power of the first reflected signal, ${P}_t\varrho^4 N^3 \beta^2_{\mathrm{r}} g^2 \approx 10^{-6} {P}_t\varrho^2 N g^2$.  The transmit power constraint of the active RIS in (\ref{vbxz}) can be simplified as 
		\begin{equation} \label{ffgbs}
			{P}_t\varrho^2 N g^2\le P_{\rm{RIS}}.
		\end{equation}	
		Similarly, the thermal noise introduced by the first reflected signal is much smaller than that introduced by  the second reflected signal, i.e., $\varrho ^4\sigma^2 N^3 g^2 \beta_{\mathrm{r}}^2 \ll 2\varrho^2 \sigma^2 N g^2$.
		
		In the active RIS systems, we need to further discuss how to allocate the transmit power budget between the radar and the active RIS with a fixed total power budget, i.e., $P_{\rm{RIS}}+P_t \le C$. Note that  it is hard to derive the close-form solution of the radar SINR, we approximate the radar SINR by ignoring the term $g_1$. In that way, the radar SINR can be approximated as
		\begin{align} \label{nmhv}
			{\rm{ SINR}}&\approx \frac{ {P}_t \varrho^4 f^4}{\sigma^2_{\rm{r}}+\varrho ^4\sigma^2 f^2 N \beta_{\mathrm{r}}+2\varrho ^2\sigma^2 N g^2} \notag \\
			&\approx\frac{ {P}_t \varrho^4 N^4 g^4 \beta_{\mathrm{r}}^2}{\sigma^2_{\rm{r}}+2\varrho^2 \sigma^2 N g^2}\triangleq{\rm{ SINR}}_{\rm{app}}.
		\end{align}
		
		The transmit power allocation problem can be formulated as follows:
		\begin{subequations}\label{optimixbgze}
			\begin{align}
				\mathop {\max }\limits_{{{P_t, P_{\rm{RIS}},\varrho} } } \quad  &	{\rm{ SINR}}_{\rm{app}} \label{bfgvbv} 
				\\
				\textrm{s.t.}\qquad &(\ref{ffgbs}), (\ref{dbs}), \label{dbgdfbs}\\
				&P_{\rm{RIS}}+P_t \le C. \label{vbxzb}
			\end{align}
		\end{subequations}
		
		\itshape \textbf{Lemma 6:} \upshape If ${P}_t a_{\rm{RIS}} N g^2\ge C$, the optimal transmit power of the active RIS $P_{\rm{RIS}}^{*} $ satisfies $P_{\rm{RIS}}^{*} \rightarrow C$. Otherwise, the optimal transmit power allocation should follow $(C-P_{\rm{RIS}}^{*})a_{\rm{RIS}} N g^2=P_{\rm{RIS}}^{*}$. 
		
		\itshape \textbf{Proof:} \upshape If ${P}_t a_{\rm{RIS}} N g^2\ge C$, it means that $\varrho^2< a_{\rm{RIS}}$ and the transmit power constraint of the active RIS (\ref{ffgbs}) is tight. By substituting (\ref{ffgbs}) into (\ref{bfgvbv}), we have 
		\begin{equation} \label{vfdz}
			\begin{aligned}
				{\rm{ SINR}}_{\rm{app}}&=\frac{ {P}_t \varrho^4 N^4 g^4 \beta_{\mathrm{r}}^2}{\sigma^2_{\rm{r}}+2\varrho^2 \sigma^2 N g^2} 
				=\frac{P_{\rm{RIS}}^2 N^2 \beta_{\mathrm{r}}^2 }{{P}_t\sigma^2_{\rm{r}}+2 \sigma^2 P_{\rm{RIS}}}, 
			\end{aligned}
		\end{equation} 
		which indicates that it is better to allocate the transmit power to the active RIS rather than the radar. 
		
		
		If ${P}_t a_{\rm{RIS}} N g^2<C$, allocating all the transmit power to the active RIS becomes infeasible owing to the limited power amplification gain $\varrho^2=a_{\rm{RIS}}$. The optimal transmit power allocation can be achieved when both (\ref{ffgbs}) and (\ref{dbs}) are tight. If more power budget is allocated to the active RIS, i.e.,  $P_{\rm{RIS}}> P_{\rm{RIS}}^{*}$, constraint (\ref{ffgbs}) and equation (\ref{vfdz}) no longer hold with equality, while the radar SINR decreases due to the decrease of $P_t$.  $\hfill\blacksquare$
		
		\itshape \textbf{Remarks:} \upshape Drawing on the insights from Lemma 6, we underscore the pivotal role of fully exploiting the power amplification potential of the active RIS. The rationale behind this is that the active RIS can amplify the signal twice while only consuming the transmit power for the first reflected signal. To be specific, the power amplification gain $\varrho^2$ exhibits a linear relationship with $P_{\rm{RIS}}$, given the negligible signal power of the second reflected signal. Consequently, the power of the received signal is a quadratic function of the power amplification gain $\varrho^2$ and a linear function of $P_t$, thereby warranting a more substantial allocation of transmit power to the active RIS.

		\subsection{Active RIS Deployment Strategy for a Moderate Number of Reflecting Elements}
		For a moderate number of reflecting elements, the transmit power constraint of the active RIS and the radar SINR can be approximated as indicated in (\ref{ffgbs}) and (\ref{nmhv}), respectively. 
		
		If ${P}_t a_{\rm{RIS}} N g^2\ge P_{\rm{RIS}}$, it follows that $\varrho^2< a_{\rm{RIS}}$ and the transmit power constraint of the active RIS is tight. In such a scenario, ${\rm{ SINR}}_{\rm{app}}$ is a monotonically increasing function of the path loss of the RIS-target-RIS link $\beta_{\mathrm{r}}^2$ and irrelevant to the path loss between the DFRC BS and the active RIS $g$ according to equation (\ref{vfdz}). Thus, it is better to deploy the active RIS near the target with limited transmit power. 
		
		If ${P}_t a_{\rm{RIS}} N g^2< P_{\rm{RIS}}$, we have $\varrho^2= a_{\rm{RIS}}$. However, it is still hard to discuss the optimal deployment strategy w.r.t. ${\rm{ SINR}}_{\rm{app}}$. For a constrained power amplification and a moderate number of reflecting elements, it can be verified that $2a_{\rm{RIS}} \sigma^2 N g^2 \ll \sigma^2_{\rm{r}}$, indicating that the impact of the distance on the denominator of the radar SINR can be considered negligible. Thus, we only discuss the radar SNR without considering the AWGN introduced by the active RIS:
		\begin{equation}
			\begin{aligned}
				\mathrm{SNR}&\triangleq\frac{ {P}_t \varrho^4 N^4 g^4 \beta_{\mathrm{r}}^2}{\sigma^2_{\rm{r}}}\\
				&\propto (d^2+x^2)^{-\alpha_{\mathrm{RIS}}}(d^2+(D-x)^2)^{-\alpha_{\mathrm{TG}}}, \\
				&\propto x^{-2\alpha_{\mathrm{RIS}}}((D-x))^{-2\alpha_{\mathrm{TG}}}, \, \textrm {when} \,  {d} \ll \mathrm{min}({x, D-x}).
			\end{aligned}
		\end{equation} 
		The path loss exponent for the BS-RIS link $\alpha_{\mathrm{RIS}}$ and RIS-Target link $\alpha_{\mathrm{TG}}$ are set to be 2.2 and 2, respectively, as indicated in equations (\ref{sthbrr}) and (\ref{grgh}).
		
		Thus, when the power amplification gain is tightly constrained by (\ref{dbs}), the radar SINR first decreases and then increases as $x$ increases and achieves its minimum value when $x^{*}=\frac{\alpha_{\mathrm{RIS}}}{\alpha_{\mathrm{RIS}}+\alpha_{\mathrm{TG}}}D$, which is similar to the findings in the passive RIS-aided communication systems\cite{9326394}.
		
		\begin{figure}[htbp]
			\vspace{-0.5cm}
			\centering
			\begin{minipage}[t]{0.48\textwidth}
				\centering
				\includegraphics[width=7cm]{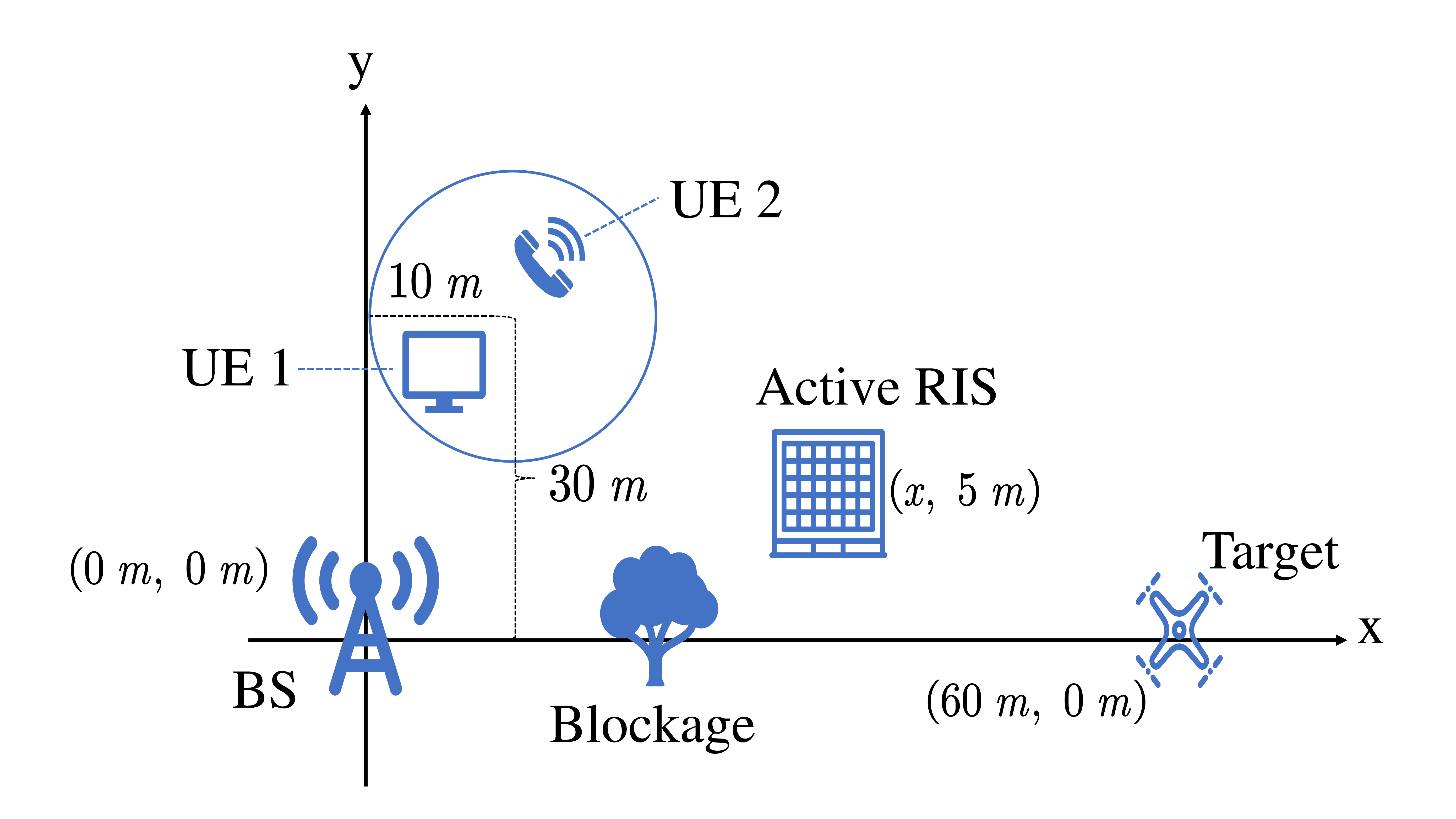}
				\caption{The simulation system setup.}
			\end{minipage}
			\begin{minipage}[t]{0.48\textwidth}
				\centering
				\includegraphics[width=7cm]{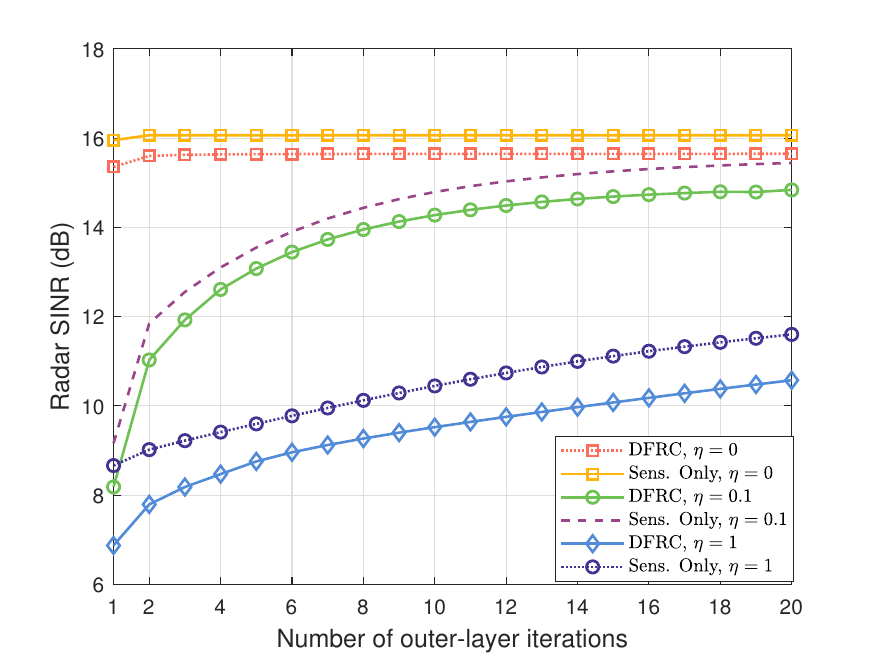}
				\caption{Convergence behaviors.}
			\end{minipage}
			\vspace{-0.5cm}
		\end{figure}	
	}

	\section{Simulation results}
	In this section, simulation results are provided to illustrate the efficiency of the proposed RIS-aided ISAC systems. It is assumed that the BS is equipped with $M=4$ antennas and communicates with two single-antenna UEs while sensing the target at the same time. Unless stated otherwise, we set the simulation parameters as follows. {\color{black}The BS, the RIS, and the target are located at (0, 0), (30 m, 5 m), and (60 m, 0), respectively. We assume that the coordinates of two UEs are generated uniformly and randomly in a circle region centered
		at (10 m, 30 m) with a radius of 10 m.}  The maximum power amplification gain $a_{\rm{RIS}}$ is assumed to be $40$ dB, and the number of reflection elements at the active RIS is $N=16$. The communication SINR threshold is set to $\xi=10$ dB, and the SINR threshold to generate the initial point of the AO algorithm is set to $\xi_2=30$ dB.  We set the carrier frequency of the system as $2.7$ GHz {\color{black} and the RCS of the target as 1 $ \mathrm{m}^2$. We also assume that the SI coefficient is given by  $\eta=0.1$\cite{SI_zhenkun}}, the channel bandwidth is 10 MHz, and the noise at the DFRC BS, UEs, and the active RIS has the same noise power density of $-174$ dBm/Hz\cite{26}. In the simulation, the results are obtained by averaging over 100 independent channel realizations.
	
	To fully evaluate the active RIS-aided ISAC system, we consider the following benchmark scenario. We denote `DFRC' as the considered scenario in (\ref{optimize}). The `Sens. only' represents the system only performs the sensing function without QoS constraints. For the `Pas. RIS' scenario, we consider a passive RIS-aided ISAC system. Specifically, the thermal noise and the transmit power constraint of the RIS do not exist  {\color{black}while the unit modulus constraint of the reflecting coefficients is considered.} The corresponding problem is solved by using the simplified Algorithm 1.

	\subsection{{Convergence Behavior with Different Levels of SI }}	
	{\color{black}In this subsection, we evaluate the convergence of the proposed algorithm under different SI coefficients, as shown in Fig. 3. When the interference from the BS-RIS-BS link is perfectly eliminated in the DFRC system (i.e., $\eta=0$), there is a performance gain of approximately 4 dB compared to the case when the BS does not operate with SI cancellation, (i.e., $\eta=1$). Additionally, we observe that the proposed algorithm requires more outer-layer iterations to converge when the interference is not effectively eliminated in the DFRC system. 
		These findings reveal that without effective SI cancellation, the interference introduced by the active RIS can have a significant impact on the interference covariance matrix. Consequently, this affects the tightness of the matrix inverse approximation used in Lemma 1 during the outer-layer iterations.}
	\subsection{{Performance Evaluation of the Active RIS-aided Sensing Systems}}
	
	{\color{black}In Fig. 4, we evaluate the performance of active RIS-aided sensing systems under two scenarios: when only the power amplification constraint `$\varrho^2=a_{\text{RIS}}$' is imposed or the transmit power budget constraint is also taken into consideration. The transmit power of the BS is set to $P_{t}=1$ W, and the SI coefficient is set to $\eta=-30$ dB. As depicted in Fig. 4, when the number of reflecting elements is moderate, irrespective of whether the transmit power budget is considered, the approximation of the radar SINR ${\rm{ SINR}}_{\rm{app}}$ and radar SNR shares a similar performance of the actual value of the radar SINR. This suggests that the conclusions drawn in Section V-C and Section V-D are applicable to most cases. Moreover, we observe that the radar SINR under the constraint `$\varrho^2=a_{\text{RIS}}$' shows similar performance as both constraints are considered for $N\le 240$. This means all the reflecting elements are operating at their maximum amplification until $N\ge 240$, and then the transmit power budget of the active RIS becomes insufficient. Let us consider a large number of the reflecting elements, i.e., $N>10^3$. When the transmit power constraint is imposed, the transmit power budget of the active RIS is set to $P_{\text{RIS}}=1$ W and the optimal power amplification gain $\varrho^2$ is obtained by applying the binary search algorithm.  Notably, in both scenarios, both the upper and lower bounds of the radar SINR, as well as the radar SINR itself, scale with the number of elements $N$.}

	\subsection{{Active RIS and Passive RIS Comparison}}
	Fig. 5 compares the radar SINR performance versus the number of RIS elements. {\color{black}Both the hardware power consumption and the transmit power of the passive and active RIS-aided ISAC systems are taken into consideration.} The overall power budget of both the active RIS, denoted as $Q_{\rm{act}}$, and the passive RIS, denoted as $Q_{\rm{pas}}$, can be respectively expressed as
	\begin{equation}
		\begin{aligned}
			Q_{\rm{act}} & =P_{\rm{BS}}+P_{\rm{RIS}}+N\left(P_{\rm{SW}}+P_{\rm{DC}}\right), \\
			Q_{\rm{pas}} & =P_{\rm{BS}}^{\text {pas }}+N P_{\rm{SW}},
		\end{aligned}	
	\end{equation}
	where $P_{\rm{BS}}^{\text {pas }}$ denotes the transmit power of BS for the passive RIS-aided ISAC system, $P_{\mathrm{SW}}=-5$ dBm denotes the power consumed by the phase control in each element of the RIS and $P_{\mathrm{DC}}=-10$ dBm denotes the direct-current (DC) power consumption in each RIS element\cite{26}. In the considered scheme, an active RIS is employed with a transmit power of $P_{\rm{RIS}}=0.1$ W, while the remaining power is allocated to the BS after accounting for hardware power consumption. {\color{black}For the passive RIS-aided systems, since it consumes only a small amount of circuit power, correspondingly more transmit power is allocated to the BS to enhance the system’s performance.} As depicted in Fig. 5, the sensing performances of both active and passive RISs improve as the number of RIS elements increases.  {Furthermore, increasing the number of the RIS reflecting elements can obtain a higher performance gain than increasing the number of antennas.} Finally, when the number of RIS elements is $N=36$, the active RIS outperforms the passive RISs with $Q_{\rm{pas}}=1$ W and $Q_{\rm{pas}}=100$ W by approximately 70 dB and 50 dB, respectively. 

	\begin{figure}[h]
		\centering
		\begin{minipage}[t]{0.45\textwidth}
			\centering
			\includegraphics[width=7cm]{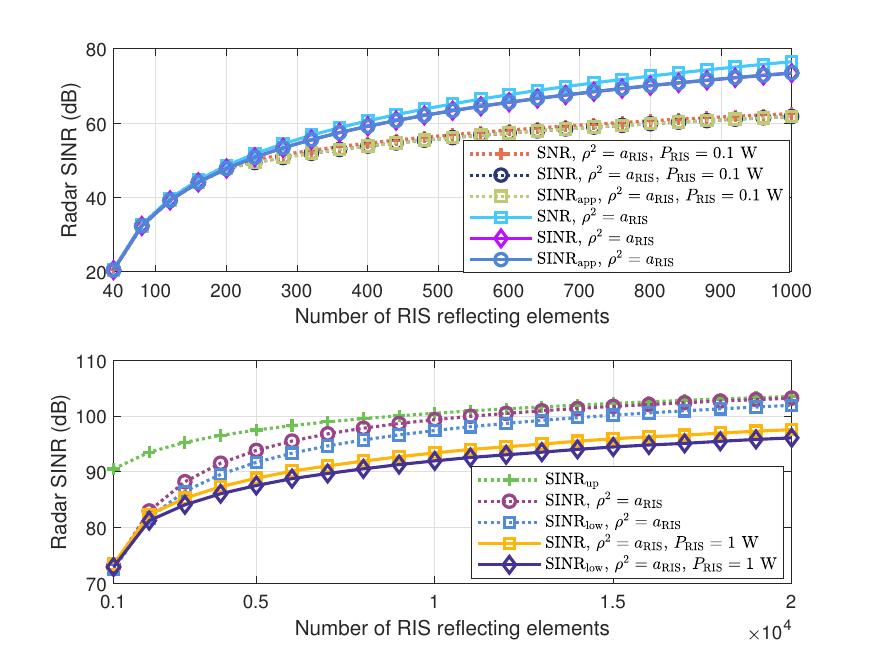}
			\caption{{\color{black}Approximate radar SINR versus the the number of RIS elements $N$.}}
		\end{minipage}	
		\begin{minipage}[t]{0.45\textwidth}
			\centering
			\includegraphics[width=7cm]{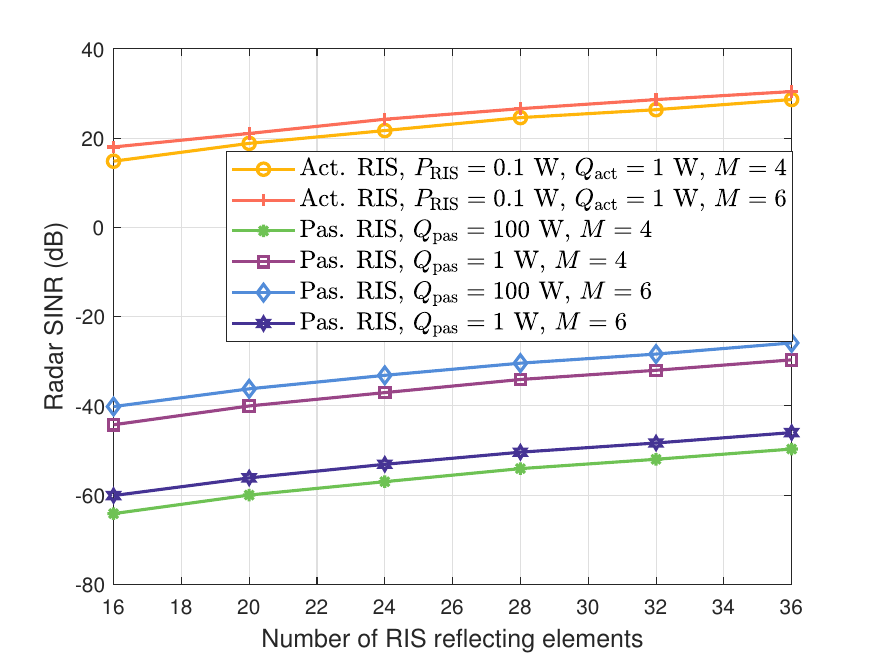}
			\caption{{\color{black}Radar SINR versus the number of RIS elements $N$.}}
		\end{minipage}	
		\begin{minipage}[t]{0.45\textwidth}
			\centering
			\includegraphics[width=7cm]{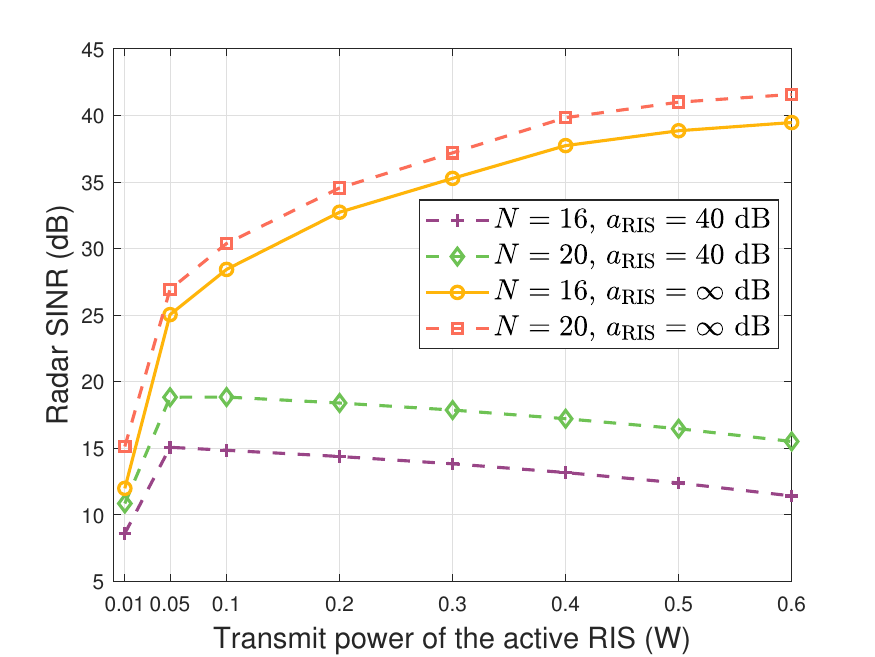}
			\caption{{\color{black}Radar SINR versus the transmit power budget of the active RIS.}}
		\end{minipage}
		\vspace{-0.5cm}
	\end{figure}
	
	\subsection{{Power Allocation Among BS and Active RIS }}
	
	In Fig. 6, we investigate the transmit power allocation between the BS and the active RIS by fixing $Q_{\rm{act}}=1$ W and varying $P_{\rm{RIS}}$ w.r.t. different numbers of the RIS elements. We first investigate the case where the amplification gain constraint is not considered. In this scenario, we find that it is advantageous to allocate more transmit power to the active RIS as discussed in Section V-C.  For the active RIS-aided ISAC systems subject to the power amplification gain constraint, the radar SINR initially improves with an increase in the RIS transmit power, but then begins to deteriorate as $P_{\rm{RIS}}$ exceeds 0.05 W. {\color{black}This behavior arises due to the fact that most of the reflecting elements operate at their maximum power amplification gain when $P_{\rm{RIS}}> 0.05$ W. Consequently, the transmit power of the active RIS does not increase as the transmit power budget of the active RIS increases. Instead, the radar SINR decreases due to the reduction in the transmit power at the DFRC BS. } Furthermore, increasing the number of reflecting elements enhances the radar SINR regardless of whether the amplification gain constraint is considered or not. 
	
	\begin{figure}[htbp]
		\centering
		\vspace{-0.2cm}
		\begin{minipage}[t]{0.45\textwidth}
			\centering
			\includegraphics[width=7cm]{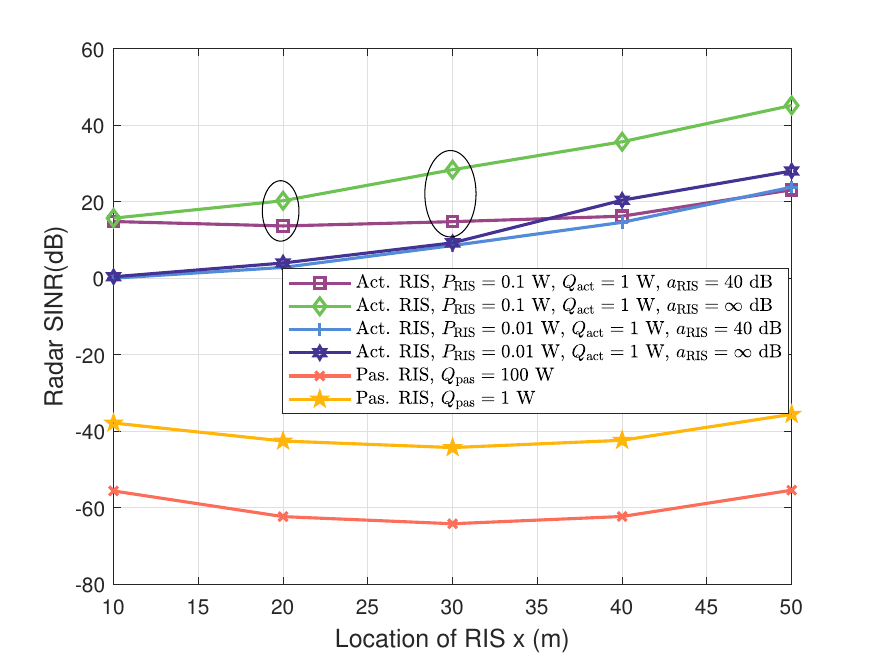}
			\caption{Radar SINR versus the location of the RIS $x$.}
		\end{minipage}
		\begin{minipage}[t]{0.45\textwidth}
			\centering
			\includegraphics[width=7cm]{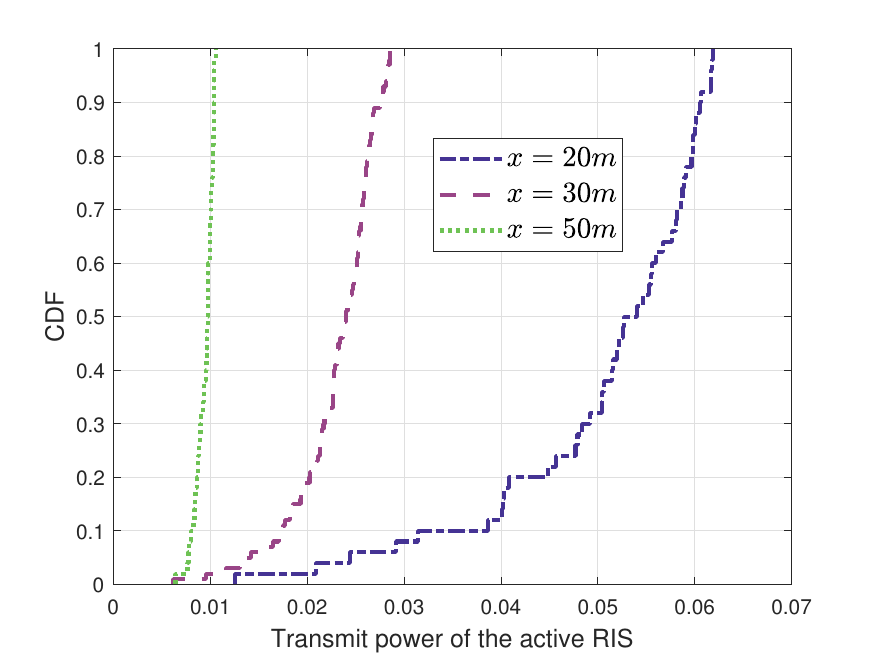}
			\caption{CDF of the transmit power of the active RIS.}
		\end{minipage}
	\end{figure}
	
	\subsection{{Deployment of the Active RIS }}	 	
	\color{black}{
		
		Fig. 7 illustrates our investigation of how the deployment of both passive and active RISs
		affects the radar SINR. As discussed in Section V-D, when the power amplification gain is not taken into account, the radar SINR is monotonically increasing w.r.t. $x$.  When the power amplification gain is considered, the radar SINR first decreases and then increases as $x$ increases and approximately achieves its minimum value when $x=30$ m, exhibiting a similar performance trend as the passive RIS. Besides, as the location of the active RIS $x$ increases, the performance gap between the active RIS with and without a constrained power amplification gain becomes more pronounced. When there is the power amplification gain constraint, as the distance between the BS and the RIS increases, the achievable transmit power on the active RIS becomes smaller. On the other hand, when not considering power amplification gain, the transmit power on the active RIS is always equal to the maximum transmission power of the active RIS. Therefore, the difference in the transmit power and the radar SINR between the two scenarios becomes larger as the distance increases.
		
		Notably, we observe that the performance gap between the active RIS with a constrained power amplification gain and different transmit power budgets (as indicated in `Act. RIS, $P_{\rm{RIS}}=0.01$ W, $a_{\rm{RIS}}=40$ dB' and `Act. RIS, $P_{\rm{RIS}}=0.1$ W, $a_{\rm{RIS}}=40$ dB') diminishes as the RIS approaches the target location. Particularly, when the RIS is situated near the target at $x=50$ m, the active RIS with a maximum transmit power of $0.01$ W demonstrates similar performance to that of the active RIS with a maximum transmit power of $0.1$ W. To elucidate this result, we examine the cumulative distribution function (CDF) of the achievable transmit power consumption of the active RIS at various locations when the RIS operates with its maximum power amplification gain of $a_{\rm{RIS}}=40$ dB and is supplied with $P_{\rm{RIS}}=0.1$ W. As depicted in Fig. 8, it is evident that in all cases, the  transmit power remains below $P_{\rm{RIS}}=0.1$ W. At $x=50$ m, the transmit power of the active RIS is approximately $0.01$ W, providing an explanation for the similar performance of the $P_{\rm{RIS}}=0.1$ W and $P_{\rm{RIS}}=0.01$ W when the active RIS is located near the target.
		
		In summary, we conclude that both the transmit power budget and the power amplification gain constraint significantly influence the deployment strategy. Generally, deploying the active RIS in proximity to the target location proves to be an optimal solution, as it yields higher radar SINR and potentially lower transmit power consumption at the active RIS.
	\color{black}

	\vspace{-0.3cm}
	\begin{figure}[htbp]
		\centering
		\begin{minipage}[t]{0.45\textwidth}
			\centering
			\includegraphics[width=7cm]{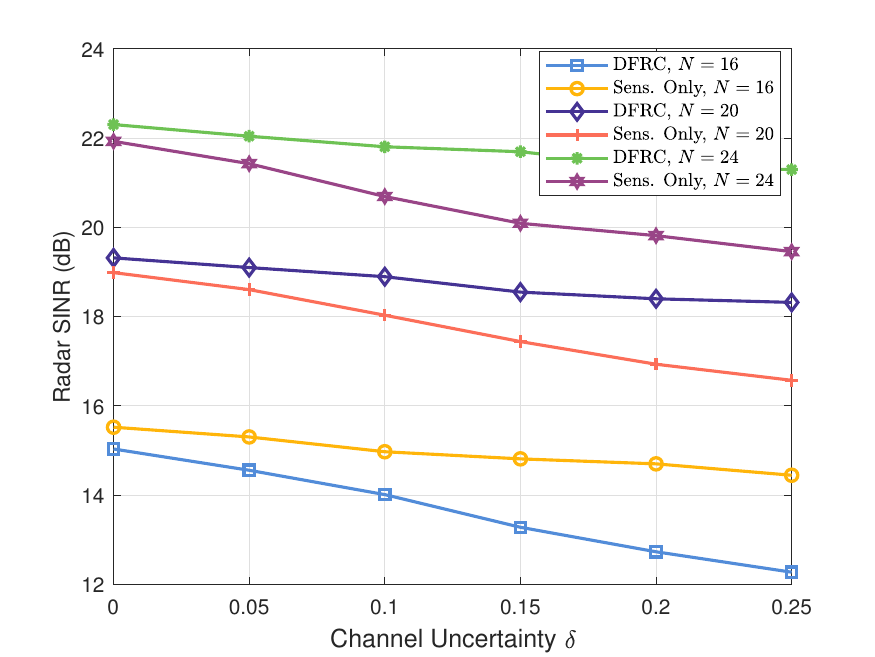}
			\caption{Radar SINR under channel uncertainty $\delta$.}
		\end{minipage}
		\begin{minipage}[t]{0.45\textwidth}
			\centering
			\includegraphics[width=7cm]{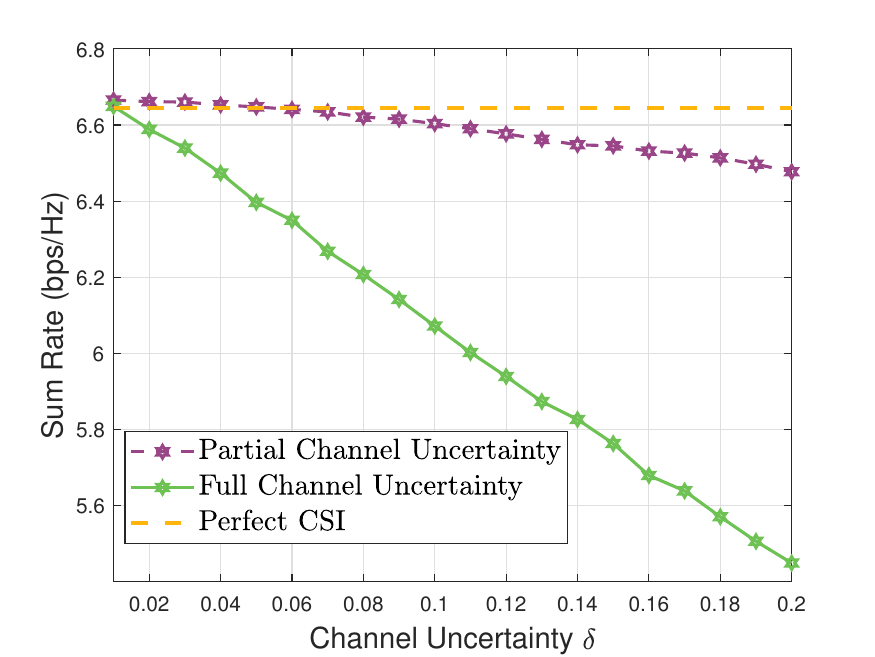}
			\caption{Sum rate under channel uncertainty $\delta$.}
		\end{minipage}
		\vspace{-0.5cm}
	\end{figure}
	
	\subsection{Radar SINR and Sum Rate with Imperfect CSI}
	{\color{black}For the passive RIS systems, it is typically sufficient to have the knowledge of the CSI of the cascaded channels for transceiver and RIS phase shift design. However, in the case of active RIS systems, the individual CSI of the BS-RIS channel $\mathbf{G}$, and the
		RIS-UE channel $\mathbf{h}_{1,k}$ are required to calculate the RIS transmit power in (\ref{RIS constrains}), the communication SINR in (\ref{SINR_k}), and the radar SINR in (\ref{SINR}) \cite{Active_Zhougui}. It is challenging to estimate the individual CSI due to the lack of signal processing capability at the active RISs. To show the robustness of the proposed scheme, we consider a benchmark containing the estimation error of $\mathbf{G}$ and $\mathbf{h}_{1,k}$. The bounded CSI error model is adopted and given by 
		\begin{equation} \label{vsd}
			\begin{aligned}
				\mathbf{G}=\widehat{\mathbf{G}}+\triangle \mathbf{G} \quad \textrm{and} \quad \mathbf{h}_{1,k}=\widehat{\mathbf{h}}_{1,k}+\triangle \mathbf{h}_{1,k}.
			\end{aligned}
		\end{equation}
		Here, $\widehat{\mathbf{G}}$ and $\widehat{\mathbf{h}}_{1,k}$ represent the estimated CSI known at the BS, while $\triangle \mathbf{G}$ and $\triangle \mathbf{h}_{1,k}$ represent the unknown estimation errors satisfying the conditions $\left\|\triangle \mathbf{G}\right\|_{F} \leq \delta \left\| \mathbf{G}\right\|_{F}$ and $\left\|\triangle \mathbf{h}_{1,k}\right\|_{2} \leq \delta \left\| \mathbf{h}_{1,k}\right\|_{2}$, respectively. $\delta \in[0,1)$ accounts for the relative amount of CSI uncertainties.
		

		As shown in Fig. 9, the radar SINR is monotonically decreasing as the level of channel uncertainty increases. When compared to the `Sens. only' scheme, the DFRC scheme experiences a more pronounced performance deterioration due to the allocation of less power towards the active RIS. With a reduced power reception at the active RIS, the impact of channel uncertainty on the radar SINR becomes more significant. 
		
		It is important to note that, in the presence of channel uncertainty, the tightness of the communication QoS constraint may be affected. To further investigate this issue, we present the achievable sum rate as a function of channel uncertainty in Fig. 10. Specifically, we compare the sum rate under partial channel uncertainty, which was presented in (\ref{vsd}), with the case of full channel uncertainty, where the bounded error model is also applied to the direct link between the BS and the UE $\mathbf{h}_{2,k}$ \cite{53}. It is observed that when the channel between the BS and UE is also imperfect, there is a significant decrease in the sum rate of the two communicating UEs. These findings reveal that, under equal path loss exponents for the direct link and the cascaded channel, considering the amplification gain constraint of the active RIS, the direct channel holds a dominant influence on the cascaded RIS link.}
	
	\begin{center}
		\begin{figure}
			\centering		\includegraphics[width=3in]{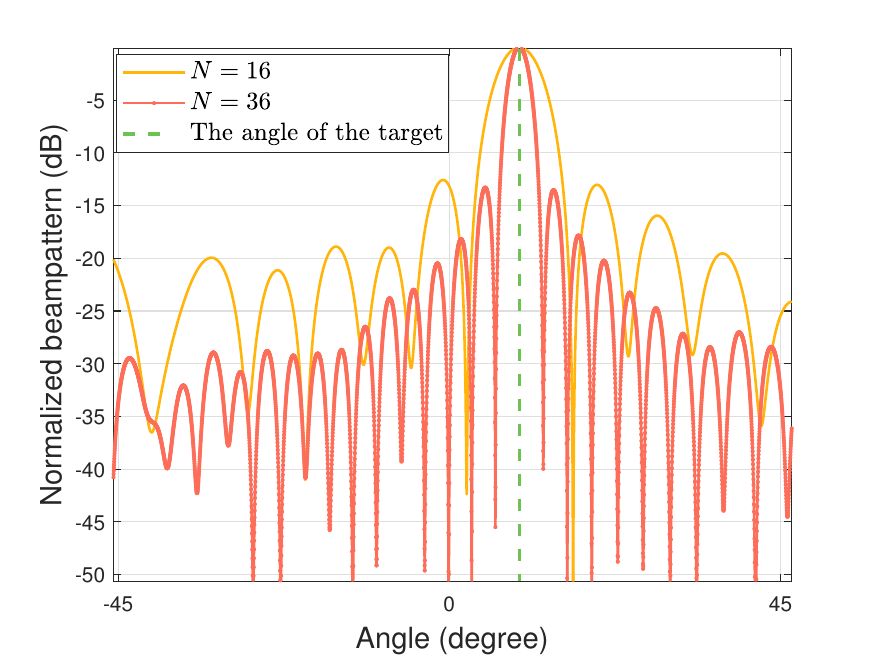}
			\		\caption{Beampattern of the proposed scheme. }
			\label{fig10}
			\vspace{-0.5cm}
		\end{figure}
	\end{center} 
	\vspace{-1cm}	
	{\color{black}
		\subsection{{Beampattern of the Proposed Scheme }}
		
		Fig. \ref{fig10} presents a comparison of the normalized sensing beampattern for different numbers of reflecting elements of the active RIS.  The beampattern gain from the active RIS towards angle $\theta$ is defined as
		\begin{equation}
			\begin{aligned}
				\mathcal{P}(\theta) & \triangleq \mathbb{E}\left[\left|\mathbf{a}_3^{\mathrm{H}}(\theta) \mathbf{\Phi} \mathbf{G}\mathbf{x}\right|^{2}\right] \\
				& =\mathbf{a}_3^{\mathrm{H}}(\theta) \mathbf{\Phi} \mathbf{G}\mathbf{R}\mathbf{G}^{\mathrm{H}} \mathbf{\Phi}^{\mathrm{H}}\mathbf{a}_3(\theta) .
			\end{aligned}
		\end{equation}
		
		As observed in Fig.\ref{fig10}, increasing the number of reflecting elements in the active RIS can enhance the performance and result in a narrower beam. However, in scenarios with inaccurate estimation of the angle between the active RIS and the target, particularly in high mobility scenarios, this can also lead to a larger  performance degradation.}
	
	{\color{black}	\section{Practical issues and open problems}
		In the aforementioned sections of this paper, several ideal assumptions are adopted for the sake of simplicity. In the following, we briefly introduce several practical issues that may warrant further attention in future research.	
		
		\subsection{Active RIS Architectures and Wideband Scenarios}
		Apart from the active RIS architecture adopted in our work, several alternative active RIS architectures with lower cost/power consumption or additional functionalities have been proposed in the recent literature\cite{Active_RIS_new,STAR_Active}. In \cite{Active_RIS_new}, an alternative active RIS architecture utilizing two single passive RISs and a power amplifier was introduced, while the authors of \cite{STAR_Active} proposed a novel double-faced active (DFA)-RIS structure that may ensure a full space coverage simultaneously.
		
		Additionally, the existing literature mainly focused on the investigation of active RISs with narrowband signals, which however cannot be extended to the wideband active RIS-aided system. Since the amplitudes and phase shifts of reflected signal change with different frequencies of incident signals\cite{wideband_active_RIS}. Further research is needed to explore the different models and performance characteristics of the active RIS-aided ISAC systems in wideband scenarios.
		
		\subsection{Advanced Signal Processing Techniques}
		Since the active RIS does not contain a complete radio frequency (RF) chain or any sensors, there are several challenging issues related to signal processing at the DFRC BS. For instance, accurate CSI acquisition of individual active RIS-aided channels at the BS is more challenging than the passive RIS-aided systems.  Furthermore, effectively mitigating interference and establishing full duplex technologies in active RIS-aided ISAC systems presents intriguing technical hurdles\cite{Full-duplex}. Additionally, dynamically configuring the reflecting coefficients of the active RIS to accurately track moving targets is an exciting and complex task that demands innovative solutions. }
	

	\vspace{-0.2cm}
	
	\section{Conclusion}
	In this paper, we studied an active RIS-aided ISAC system with a four-hop sensing link. Specifically, we addressed the radar SINR maximization problem by jointly optimizing the beamforming matrix at the DFRC BS and the reflecting coefficient matrix at the active RIS, while guaranteeing the transmit power constraints of the BS and the active RIS and QoS constraints of communication UEs. To tackle the optimization problem, the MM algorithm was applied to address the nonconvex radar SINR objective function, and the resulting quartic MSR problem was solved by developing an SDR-based approach. {\color{black}In the following, we analyzed a simplified active RIS-aided sensing system. With an increasing number of reflecting elements $N$, the radar SINR exhibits a scaling order of $N$. When the number of reflecting elements is moderate, the transmit power of the active RIS should be chosen carefully to achieve a high radar SINR while operating within the same power budget. Furthermore, in the proposed scenario, deploying the active RIS in close proximity to the intended target would be a more optimal strategy. Our simulation results verified the aforementioned results and revealed that the implementation of an active RIS can significantly alleviate the detrimental multiplicative fading impact within a four-hop sensing link compared to a passive RIS configuration.} Overall, our study offered insights into the joint design of an active RIS-aided ISAC system with a four-hop sensing link and highlighted the potential advantages of integrating the active RIS in future wireless communication and sensing systems.

	\vspace{-0.2cm}
	
	\numberwithin{equation}{section}
	\begin{appendices}	
		\section {Proof for Lemma 1}
		Although the original problem is not a convex function of ${\mathbf{W}}$ or $\mathbf{\Phi}$, we define $\mathbf{X}=\mathbf{BW}$, and it can be verified that the expression $g(\mathbf{X},\mathbf{X}^{*},\mathbf{J})=\operatorname{Tr}\left(\mathbf{X}^{\rm{H}} \mathbf{J}^{-1}\mathbf{X}\right)$ is jointly convex of $\{\mathbf{X},\mathbf{J}\}$.
		For the convex function $g(\mathbf{X},\mathbf{X}^{*},\mathbf{J})$, linearlizing $ g $ at $ \mathbf{X}=\mathbf{X}_i $ yields the following inequality:
		\begin{equation} \label{nysb}
			\begin{aligned}
				&g(\mathbf{X},\mathbf{X}^{*},\mathbf{J}) \geq  \, g(\mathbf{X}_i,\mathbf{X}^{*}_i,\mathbf{J}_i)+\operatorname{Tr}\left(\left(\left.\frac{\partial g}{\partial \mathbf{X}}\right|_{\mathbf{X}_i}\right)^{\mathrm{T}}(\mathbf{X}-\mathbf{X}_i)\right) \\
				&+\operatorname{Tr}\left(\left(\left.\frac{\partial g}{\partial \mathbf{X}^{*}}\right|_{\mathbf{X}^{*}_i}\right)^{\mathrm{T}}(\mathbf{X}^{*}-\mathbf{X}^{*}_i)\right)\\
				&+\operatorname{Tr}\left(\left(\left.\frac{\partial g}{\partial \mathbf{J}}\right|_{\mathbf{J}_i}\right)^{\mathrm{T}}(\mathbf{J}-\mathbf{J}_i)\right).	
			\end{aligned}
		\end{equation}
		According to \cite{zhang2017matrix}, we obtain the following first-order derivatives:
		\begin{equation} \label{btrs}
			\frac{\partial}{\partial \mathbf{X}} \operatorname{Tr}(\mathbf{J}^{-1}\mathbf{ X X}^{\rm{H}})=\mathbf{J}^{\rm{-T}} \mathbf{X}^{*},
		\end{equation}
		\begin{equation} \label{sns}
			\frac{\partial}{\partial \mathbf{X}^*} \operatorname{Tr}\left(\mathbf{X} \mathbf{X}^{\rm{H}} \mathbf{J}^{-1}\right)=\mathbf{J}^{-1} \mathbf{X}, 
		\end{equation}
		\begin{equation} \label{ntr}
			\frac{\partial}{\partial \mathbf{J}} \operatorname{Tr}\left(\mathbf{X}^{\rm{H}} \mathbf{J}^{-1}\mathbf{X}\right)=-\left(\mathbf{J}^{-1} \mathbf{X} \mathbf{X}^{\rm{H}} \mathbf{J}^{-1}\right)^{\rm{T}}.
		\end{equation}
		By substituting  (\ref{btrs})-(\ref{ntr}) into (\ref{nysb}), the lower bound can be obtained as
		\begin{equation} 
			\begin{aligned}
				&\operatorname{Tr}\left(\mathbf{X}^{\rm{H}} \mathbf{J}^{-1}\mathbf{X}\right) 
				\geq 2\operatorname{Re}(\operatorname{Tr}\left(\mathbf{X}^{\rm{H}}_i\mathbf{J}^{-1}_i\mathbf{X}\right)) \\
				&-\operatorname{Tr}\left(\mathbf{J}^{-1}_i\mathbf{X}_i \mathbf{X}_i^{\rm{H}}\mathbf{J}^{-1}_i\mathbf{J}\right). 
			\end{aligned}
		\end{equation}
		By using the above derivations, we prove that the surrogate function satisfies the lower bound Condition 3 in Section IV-A, and also satisfies Condition 1 and Condition 2. Hence, the proof is completed.
		
		\section {Proof for Lemma 2}
		Given that $\operatorname{Rank}(\tilde{\mathbf{H}})=K$, matrix $\tilde{\mathbf{H}}^{\rm{H}}\tilde{\mathbf{H}}$ is invertible. The feasible solution is attained through constructing the beamforming matrix $\mathbf{W}^{\star}$ as 
		\begin{equation} \mathbf{W}^{\star}=\sqrt{\xi}\operatorname{Diag}{([\tilde{{d}}_1,\cdots,\tilde{{d}}_K])}^{\frac{1}{2}}\;\tilde{\mathbf{H}}(\tilde{\mathbf{H}}^{\rm{H}}\tilde{\mathbf{H}})^{-1}. 
		\end{equation} 
		By using this type of zero-forcing (ZF) beamforming method, it can be verified that $|\mathbf{h}_k^{\rm{H}}\mathbf{w}^{\star}_k|^2=\|\mathbf{h}_k^{\rm{H}}\mathbf{W}^{\star}\|_2^2=\xi d_k$, thus the QoS constraints (\ref{communication constrain7}) are satisfied. The transmit power constraint of the BS and the active RIS can be respectively given by
		\begin{equation} \xi \operatorname{Tr}(\operatorname{Diag}{([\tilde{{d}}_1,\cdots,\tilde{{d}}_K])}(\tilde{\mathbf{H}}^{\rm{H}}\tilde{\mathbf{H}})^{-1})\le P_{{t}}, 
		\end{equation}
		\begin{align}  \rho^4\|\mathbf{AG}\mathbf{W}^{\star}\|_F^2+\rho^2\|\mathbf{G}\mathbf{W}^{\star}\|_F^2+\rho^22\sigma^2 \\ \notag
			+\rho^4\sigma^2\operatorname{Tr}(\mathbf{AA}^{\rm{H}})	\le P_{\rm{RIS}}. 
		\end{align} Thus, we conclude that if there exists $\rho$ that satisfies the conditions in Section IV-B, Problem (\ref{optimize for precoding7}) is feasible. Hence, the proof is completed.

	\end{appendices}

	\
	\vspace{-0.5cm}
	\bibliographystyle{IEEEtran}
	\bibliography{IEEEabrv,mylib}

	\section*{Biography Section}
	\begin{IEEEbiography}[{\includegraphics[width=1in,height=1.25in,clip,keepaspectratio]{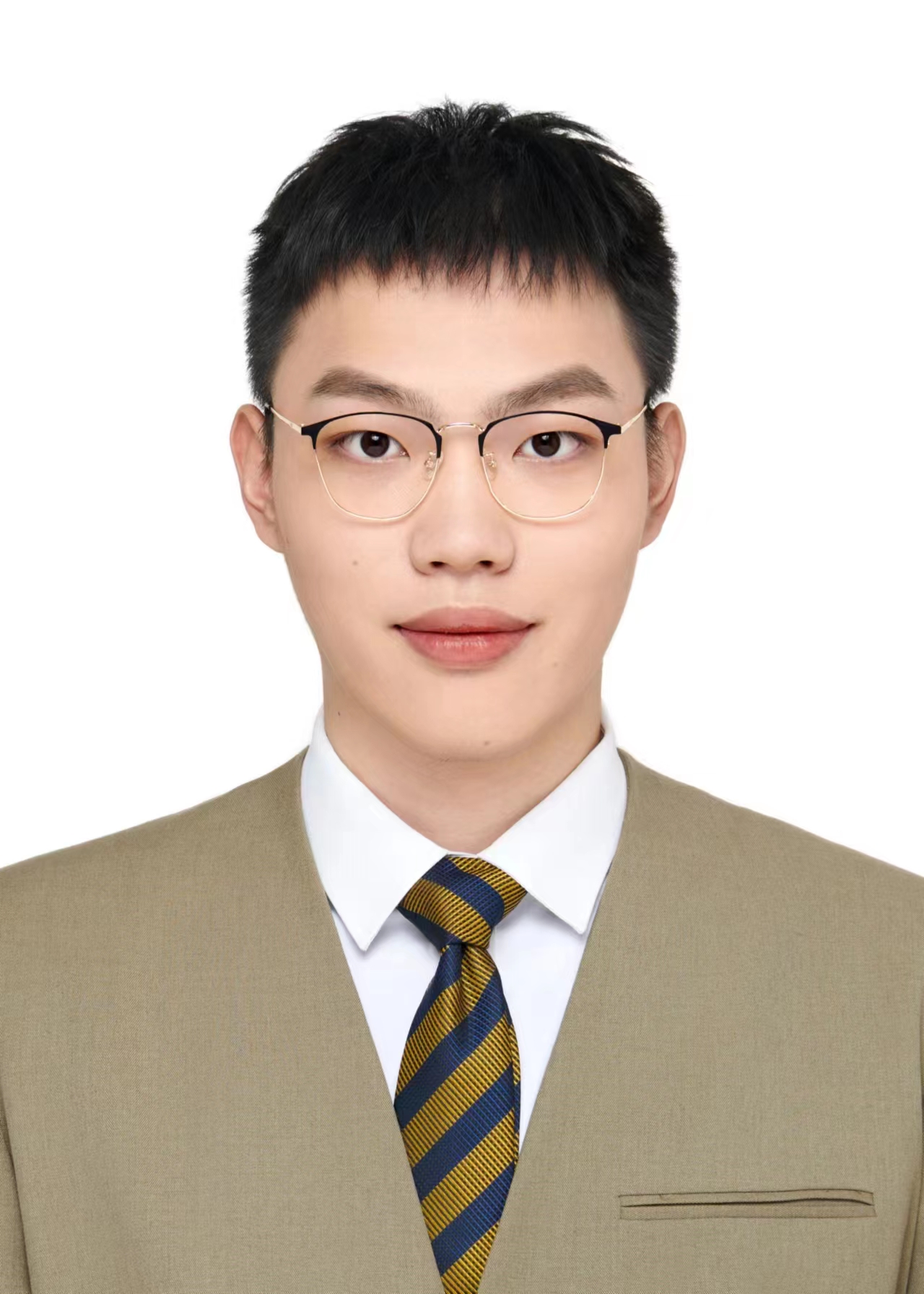}}]
		{Zhiyuan Yu} received the B.S. degree from the School of Information Science and Engineering, Southeast University, Nanjing, China, in 2023, where he is currently working toward the Master degree. His research interests include reconfigurable intelligent surfaces (RIS), integrated sensing and communication (ISAC).
	\end{IEEEbiography}

	\begin{IEEEbiography}[{\includegraphics[width=1in,clip,keepaspectratio]{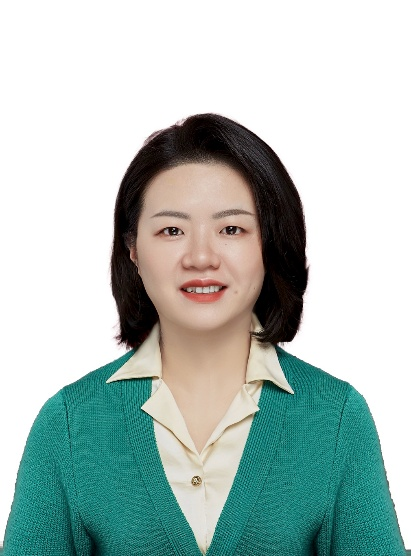}}]
		{Hong Ren}  (Member, IEEE) received the B.S. degree in electrical engineering from Southwest Jiaotong University, Chengdu, China, in 2011, and the M.S. and Ph.D. degrees in electrical engineering from Southeast University, Nanjing, China, in 2014 and 2018, respectively. From 2016 to 2018, she was a Visiting Student with the School of Electronics and Computer Science, University of Southampton, U.K. From 2018 to 2020, she was a Post-Doctoral Scholar with Queen Mary University of London, U.K. She is currently an associate professor with Southeast University. Her research interests lie in the areas of communication and signal processing, including ultra-low latency and high reliable communications, Massive MIMO and machine learning. She received the IEEE ComSoc Leonard G. Abraham Prize in 2022, IEEE WCSP 2022 Best Paper Award.
	\end{IEEEbiography}
	
	\begin{IEEEbiography}[{\includegraphics[width=1in,clip,keepaspectratio]{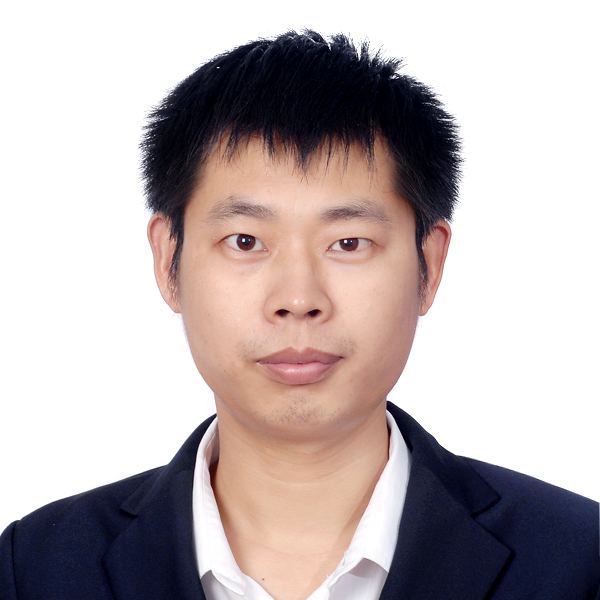}}]
		{Cunhua Pan}  (Senior Member, IEEE) 
		received the B.S. and Ph.D. degrees from the School of Information Science and Engineering, Southeast University, Nanjing, China, in 2010 and 2015, respectively. From 2015 to 2016, he was a Research Associate at the University of Kent, U.K. He held a post-doctoral position at Queen Mary University of London, U.K., from 2016 and 2019. From 2019 to 2021, he was a Lecturer in the same university. From 2021, he is a full professor in Southeast University. 
		
		His research interests mainly include  reconfigurable intelligent surfaces (RIS), intelligent reflection surface (IRS), ultra-reliable low latency communication (URLLC) , machine learning, UAV, Internet of Things, and mobile edge computing. He has published over 120 IEEE journal papers. He is currently an Editor of IEEE Transactions on Vehicular Technology, IEEE Wireless Communication Letters, IEEE Communications Letters and IEEE ACCESS. He serves as the guest editor for IEEE Journal on Selected Areas in Communications on the special issue on xURLLC in 6G: Next Generation Ultra-Reliable and Low-Latency Communications. He also serves as a leading guest editor of IEEE Journal of Selected Topics in Signal Processing (JSTSP)  Special Issue on Advanced Signal Processing for Reconfigurable Intelligent Surface-aided 6G Networks, leading guest editor of IEEE Vehicular Technology Magazine on the special issue on Backscatter and Reconfigurable Intelligent Surface Empowered Wireless Communications in 6G, leading guest editor of IEEE Open Journal of Vehicular Technology on the special issue of Reconfigurable Intelligent Surface Empowered Wireless Communications in 6G and Beyond, and leading guest editor of IEEE ACCESS Special Issue on Reconfigurable Intelligent Surface Aided Communications for 6G and Beyond. He is Workshop organizer in IEEE ICCC 2021 on the topic of Reconfigurable Intelligent Surfaces for Next Generation Wireless Communications (RIS for 6G Networks), and workshop organizer in IEEE Globecom 2021 on the topic of Reconfigurable Intelligent Surfaces for future wireless communications. He is currently the Workshops and Symposia officer for Reconfigurable Intelligent Surfaces Emerging Technology Initiative. He is workshop chair for IEEE WCNC 2024, and TPC co-chair for IEEE ICCT 2022. He serves as a TPC member for numerous conferences, such as ICC and GLOBECOM, and the Student Travel Grant Chair for ICC 2019.  He received the  IEEE ComSoc Leonard G. Abraham Prize in 2022, IEEE ComSoc Asia-Pacific Outstanding Young Researcher Award, 2022.
	\end{IEEEbiography}

	\begin{IEEEbiography}[{\includegraphics[width=1in,clip,keepaspectratio]{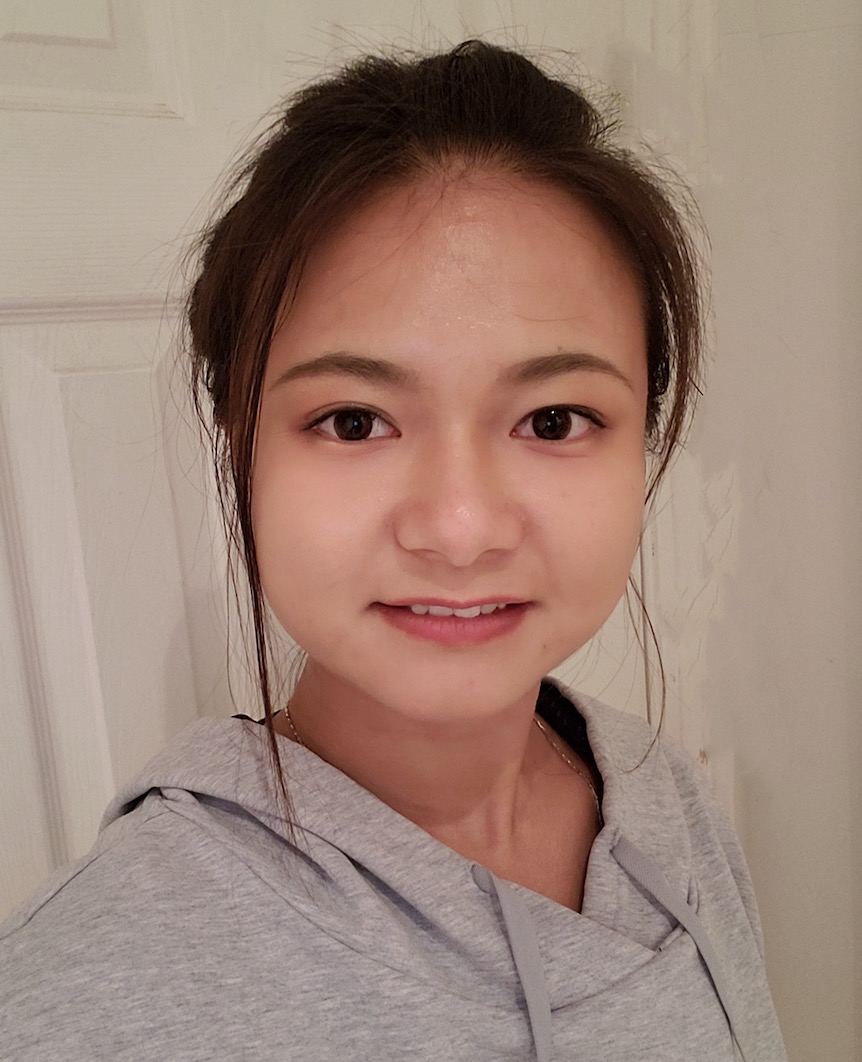}}]
		{Gui Zhou} (Member, IEEE) received the {B.S.} and {M.E.} degrees from the School of Information and Electronics, Beijing Institute of Technology, Beijing, China, in 2015 and 2019, respectively, and the {Ph.D.} degree from the School of electronic Engineering and Computer Science, Queen Mary University of London, U.K. in 2022. She is currently a Humboldt Post-Doctoral Research Fellow with the Institute for Digital Communications, Friedrich-Alexander-University Erlangen-N\"{u}rnberg (FAU), Erlangen, Germany. Her major research interests include reconfigurable intelligent surfaces (RIS) and array signal processing.
	\end{IEEEbiography}
	
	\begin{IEEEbiography}[{\includegraphics[width=1in,clip,keepaspectratio]{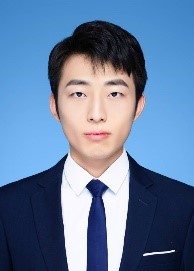}}]
		{Boshi Wang} received the B.S. degree from the School of Information Science and Engineering, Southeast University, Nanjing, China, in 2023, where he is currently working toward the Master degree. His research interests include reconfigurable intelligent surfaces (RIS).
	\end{IEEEbiography}

	\begin{IEEEbiography}[{\includegraphics[width=1in,clip,keepaspectratio]{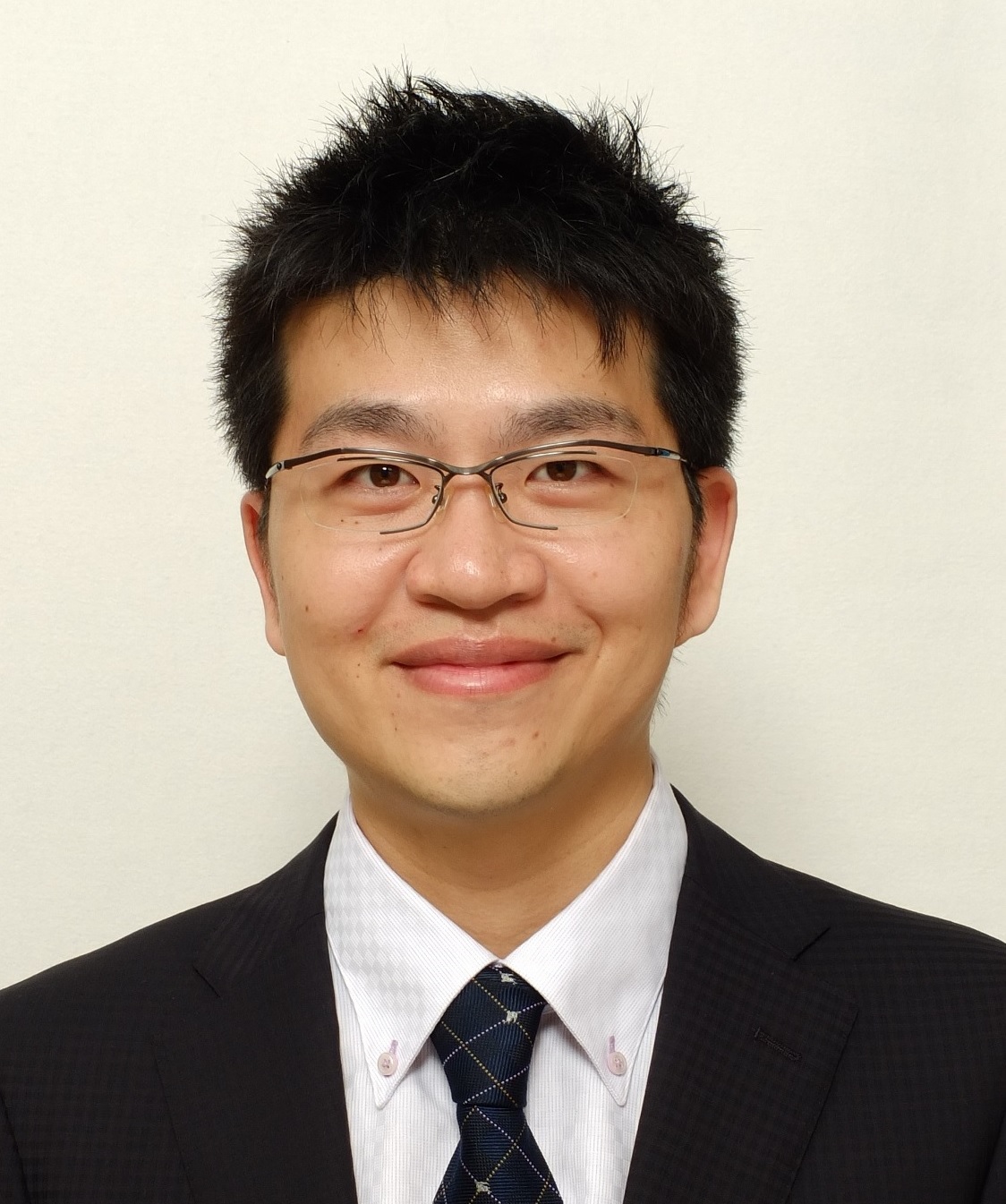}}]
		{Mianxiong Dong}  received B.S., M.S. and Ph.D. in Computer Science and Engineering from The University of Aizu, Japan. He is the Vice President and Professor of Muroran Institute of Technology, Japan. He was a JSPS Research Fellow with School of Computer Science and Engineering, The University of Aizu, Japan and was a visiting scholar with BBCR group at the University of Waterloo, Canada supported by JSPS Excellent Young Researcher Overseas Visit Program from April 2010 to August 2011. Dr. Dong was selected as a Foreigner Research Fellow (a total of 3 recipients all over Japan) by NEC C\&C Foundation in 2011. He is the recipient of The 12th IEEE ComSoc Asia-Pacific Young
		Researcher Award 2017, Funai Research Award 2018, NISTEP Researcher 2018 (one of only 11 people in Japan) in recognition of significant contributions in science and technology, The Young Scientists’ Award from MEXT in 2021, SUEMATSU-Yasuharu Award from IEICE in 2021, IEEE
		TCSC Middle Career Award in 2021. He is Clarivate Analytics 2019, 2021, 2022 Highly Cited Researcher (Web of Science) and Foreign Fellow of EAJ.
		
	\end{IEEEbiography}

	\begin{IEEEbiography}[{\includegraphics[width=1in,clip,keepaspectratio]{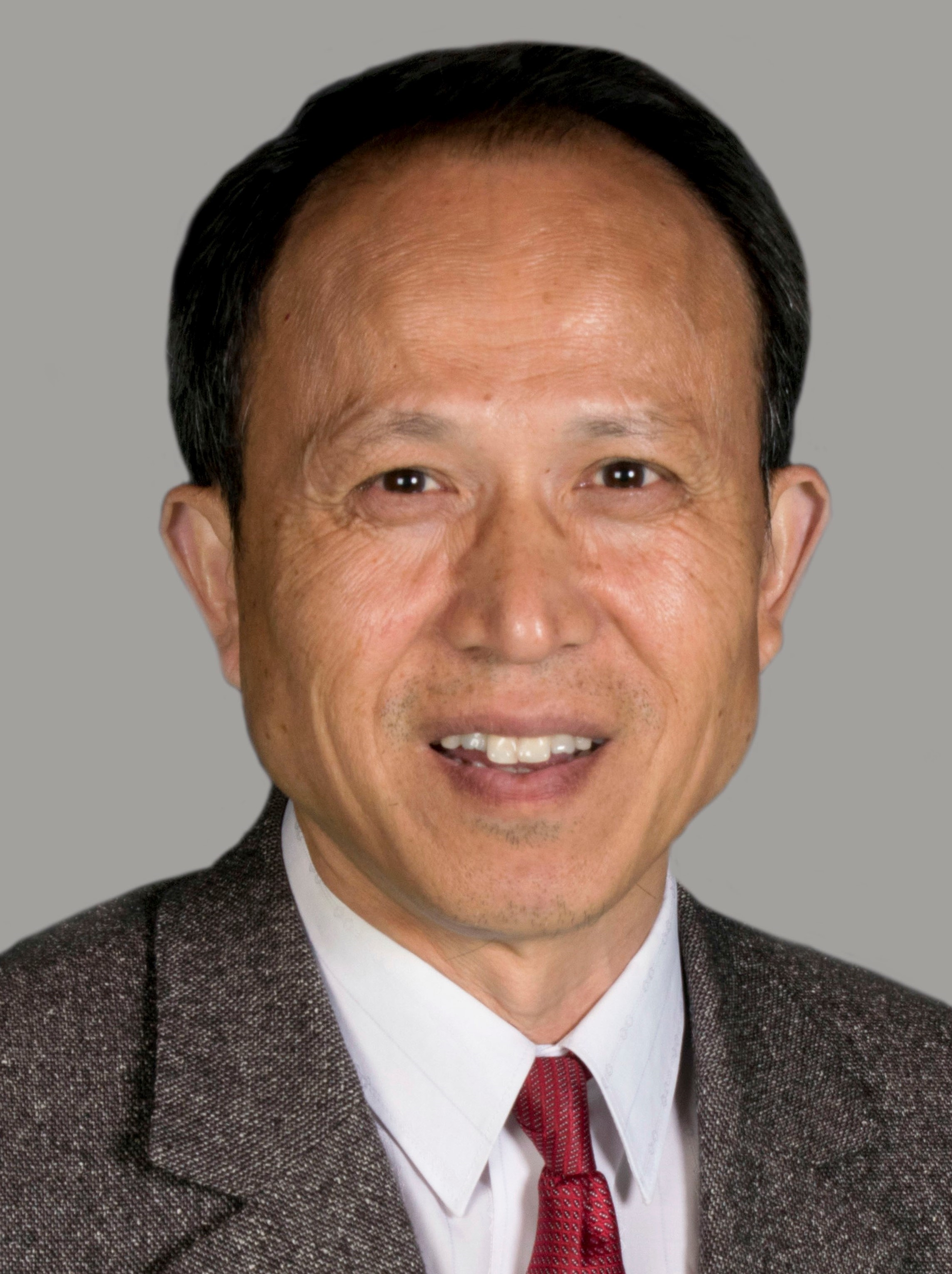}}]
		{Jiangzhou Wang}
		(Fellow, IEEE) is a Professor with the University of Kent, U.K. He has published more than 400 papers and four books. His research focuses on mobile communications. He was a recipient of the 2022 IEEE Communications Society Leonard G. Abraham Prize and IEEE Globecom2012 Best Paper Award. He was the Technical Program Chair of the 2019 IEEE International Conference on Communications (ICC2019), Shanghai, Executive Chair of the IEEE ICC2015, London, and Technical Program Chair of the IEEE WCNC2013. He is/was the editor of a number of international journals, including IEEE Transactions on Communications from 1998 to 2013. Professor Wang is a Fellow of the Royal Academy of Engineering, U.K., Fellow of the IEEE, and Fellow of the IET.
	\end{IEEEbiography}

\end{document}